\documentclass[11pt,a4paper]{article}
\usepackage{jheppub}
\usepackage{slashed}

\usepackage[utf8]{inputenc}
\usepackage{amsmath}
\usepackage{amsfonts}
\usepackage{amssymb}
\usepackage{braket}
\usepackage{multirow}
\usepackage{graphicx}

\newcommand{\beq}{\begin{equation}}
\newcommand{\eeq}{\end{equation}}
\newcommand{\bea}{\begin{eqnarray}}
\newcommand{\eea}{\end{eqnarray}}

\def\ra{\rightarrow}

\def\ld{\lambda}
\def\f{\frac}

\def\L{\left(}
\def\R{\right)}

\def\ra{\rightarrow}

\def\ld{\lambda}

\def\su5u1{SU(5) \times U(1)}
\def\fsu5u1{SU(5) \times U(1)'}
\def\so10{SO(10)}
\def\sq20{SO(10) \times SO(10)}

\usepackage{amssymb}

\title{New Insights of Electroweak Phase Transition in NMSSM}

\author[a]{Weicong Huang,}
\author[b,c]{Zhaofeng Kang,}
\author[a]{Jing Shu,}
\author[a]{Peiwen Wu,}
\author[a]{Jin Min Yang}

\affiliation[a]{State Key Laboratory of Theoretical Physics, Institute of
Theoretical Physics, Chinese Academy of Sciences,Beijing 100190, People's Republic of China.\footnote{jshu@itp.ac.cn}}
\affiliation[b]{Center for High-Energy Physics, Peking University, Beijing, 100871, P. R. China.}
\affiliation[c]{School of Physics, Korea Institute for Advanced Study, Seoul 130-722, Korea.}

\abstract{We perform a detailed semi-analytical analysis of the electroweak phase transition (EWPT) property in NMSSM, which serves as a good benchmark model in which the 126 GeV Higgs mixes with a singlet. In this case, a strongly first order electroweak phase transition (SFOEWPT) is achieved by the tree-level effects and the phase transition strength $\gamma_c$ is determined by the vacua energy gap at $T=0$. We make an anatomy of the energy gap at both tree-level and loop-level and extract out a dimensionless phase transition parameter $R_\kappa \equiv 4 \kappa v_s / A_\kappa$, which can replace $A_\kappa$ in the parameterization and affect the light CP odd and even Higgs spectra. We find that SFOEWPT only occurs in $R_\kappa \sim -1$ and positive $R_\kappa \lesssim \mathcal{O}(10)$, 
which in the non-PQ limit case would prefer either a relatively light CP odd or CP even Higgs boson $\sim (60, 100)$ GeV, therefore serves as a smoking gun signal and requires new search strategies at the LHC.
}

\begin{document}

\maketitle


\section{Introduction and motivations}

In the last two years, the ATLAS and CMS collaborations have
established the discovery of the long expected standard model
(SM)-like Higgs boson $h$, with a significance up to 6.1 and 6.9
$\sigma$, respectively~\cite{LHC:Higgs}. This new resonance has a relatively light mass $m_h\simeq126$ GeV, and its observed production or decay rate is close to the SM one. With more data accumulating, we would enter into the territory of precise understanding of EWSB mechanism. In an orthogonal direction, one may wonder its impacts on the weak scale cosmology, in particular the corresponding thermal property: the nature of electroweak phase transition (EWPT). It is not only a big question of early cosmology per se, it will also help us understand the origin of baryon asymmetry in a sense that a strongly first order EWPT (SFOEWPT) is required for successful electroweak baryogenesis (EWBG).

Baryogenesis has a close relation with the Higgs physics and moreover the successful baryogenesis implies a non-standard Higgs boson (For discussions on CP violation, see Ref. \cite{Shu:2013uua}). Broadly speaking, with the current LHC data on Higgs production and decay, we can specify three classes of SFOEWPT models based on their discovery potential through Higgs physics. The first class is that there is a colored or electric charged particle which couples to the 126 GeV Higgs boson. In this case, new particles which alter the Higgs production or decay through gluon fusion or di-photon decay channel will change the Higgs effective potential and potentially enhance the EWPT strength \cite{Chung:2012vg, Huang:2012wn, Katz:2014bha, Li:2014wia, AbdusSalam:2013eya, Laine:2013raa, Huo:2013fga,Laine:2012jy,Davoudiasl:2012tu}. Comprehensive studies have been carried out after the LHC data and it is found out that for a single particle, a SFOEWPT requires enhanced gluon fusion production rate and suppressed Higgs di-photon decay width \cite{Chung:2012vg, Huang:2012wn}. This problem can be cured if one introduces another particle with its loop contributions opposite to the first one while the EWPT strength is \textsl{enhanced} \cite{Huang:2012wn}. The second class is that we have a singlet scalar which couples to the Higgs but never develops a VEV \cite{Henning:2014gca}. In this case, future precision electroweak and Higgs measurements would constrain the overall kinematical renormalization of the 126 GeV Higgs induced by this model. The last class is that the extra scalar gets a VEV and mixes with the Higgs (or through a tadpole term which is essentially the same \cite{Damgaard:2013kva}) or there are multi-Higgs \cite{Dorsch:2013wja}. In this case, it is the mixing effect that changes the Higgs physics properties. Investigating its genuine features clearly is an important task.

Supersymmetry (SUSY) is a well motivated example among beyond SM models and it can also  
provide the SFOEWPT for successful EWBG. For instance, in the minimal supersymmetric SM (MSSM) the significant Higgs-stop coupling can lead to an acceptable EWPT strength in a tiny window, given a well organized stop sector~\cite{Carena:1996wj}. In light of the recent LHC Higgs discovery and stop exclusion, this window has been severely constrained and essentially ruled out~\cite{Carena:2012np, Cohen:2012zza, Curtin:2012aa} (For a remedy, see~\cite{Huang:2012wn}). In this class of model, one challenge after the Higgs discovery is to lift $m_h$ with the least fine-tuning while still accommodate the Higgs constrains. One simple extension is the NMSSM, which provides a large tree-level Higgs mass and a natural solution to the $\mu$ problem. With an extra singlet in the Higgs sector, it is conceivable SFOEWPT is still viable in NMSSM and we are curious on the phase transition patterns constrained by the current data. In this article, we have studied this problem in great detail and found out a critical parameter $R_\kappa \equiv 4 \kappa v_s / A_\kappa$ where SFOEWPT only occurs in $R_\kappa \sim -1$ and positive $R_\kappa \lesssim \mathcal{O}(10)$, which in turn would prefer a lighter CP odd or even Higgs boson.


This paper is organized as follows. In Section 2 we review the NMSSM in detail, including both its Higgs potential in zero and finite temperature. In Section 3, we first analyze the SFOEWPT in NMSSM semi-analytically through both the tree level and loop effects, and then provide the numerical results of the parameter scan which includes all the current experimental and strong electroweak phase transition conditions. We also show the corresponding particle spectra patterns, LHC observations and dark matter in section 4. Finally we conclude and give a discussion in Section 5 and some necessary details in the paper are given in the Appendices.

\section{The NMSSM at Zero and Non-zero Temperature}

As mentioned in the introduction, the NMSSM can accommodate natural
SUSY with the current data constraints and provides viable dark matter candidates, thus receives a lot of attention. It also provides a good benchmark model that the 125 GeV Higgs mixes with a singlet getting a VEV  , therefore provides tree-level cubic terms in the Higgs effective potential to enhance the strength of EWPT. In this section we review the basic formulas for the Higgs effective potential setup without and with the finite temperature corrections.

\subsection{Tree-level Higgs Potential}

All of the above eminent features of the NMSSM are traced back to
the Higgs sector, which in the $Z_3-$ invariant form, is written
as
\begin{align}
W_{Z_3}&\supset \lambda {S}{H}_u \cdot{H}_d
     + \frac{\kappa}{3} {S}^3,\cr
     -{\cal L}_{soft}&=
\left(\lambda A_{\lambda}H_{u}\cdot H_{d}S + \frac{1}{3}\kappa A_{\kappa}S^{3}+\mathrm{h.c.}\right).
\end{align}
After $S$ getting a VEV $v_s\equiv \langle S\rangle $ around the
electroweak (EW) scale, an effective $\mu$-term is generated. This
is the original motivation of the singlet extension. But as a
great bonus, the model provides a Higgs quartic term
at tree level, i.e., $\ld^2|H_u^0H_d^0|^2$, which can be significant
for a large $\ld\sim 1$ and moreover a small $\tan\beta\sim1$. As a
consequence, the tree-level mass of the SM-like Higgs boson
becomes:
\begin{align}\label{mh}
m_h^2=\L m_Z^2\cos^22\beta+\ld^2\sin^22\beta\R+\delta m_{\rm mix},
\end{align}
where $\delta m_{\rm mix}$ stands for the mixing effects on Higgs boson mass.
It can be sizable (typically a few GeV), positive or negative depending on the mass
order of the SM-like Higgs boson mass among the neutral Higgs bosons~\cite{Miller:2003ay,Kang:2012sy}. If $h_{\rm SM}$ is the lightest one ($H_1-$scenario), the effect is a reduction. In contrast, 
if $h_{\rm SM}$ is the next lightest one ($H_2-$scenario), the effect is an enhancement. After the LHC Higgs 126 GeV Higgs discovery, those spectra pattern of those two scenarios have been studied intensively~\cite{Kang:2013rj,Cao:2013gba,Agashe:2012zq,Christensen:2013dra,Beskidt:2013gia,Cheng:2013hna,Moretti:2013lya,Badziak:2013bda,Barbieri:2013hxa,Gherghetta:2012gb,King:2012tr,Bae:2012am,Gunion:2012gc,Jeong:2012ma,Cao:2012yn,Gunion:2012zd,King:2012is,Kang:2012sy}. Here we will investigate the status of EWPT in these two scenarios separately.


For later convenience, we give the complete tree-level Higgs
potential, which consists of the D-, F- and the soft SUSY breaking
terms:
\begin{eqnarray}\label{treeHiggs}
\nonumber V_{0}&=&\left|\lambda H_{u}\cdot H_{d}-\kappa S^{2}\right|^{2}+\left|\lambda S\right|^{2}(H_{d}^{\dagger}H_{d}+H_{u}^{\dagger}H_{u})\\
\nonumber &+&\frac{g^{2}}{4}(H_{u}^{\dagger}H_{u}-H_{d}^{\dagger}H_{d})^{2}+\frac{g_{2}^{2}}{2}\left|H_{d}^{\dagger}H_{u}\right|^{2}\\
\nonumber &+&m_{H_d}^{2}H_{d}^{\dagger}H_{d}+m_{H_u}^{2}H_{u}^{\dagger}H_{u}+m_{S}^{2}|S|^{2}\\
&+&\big(\lambda A_{\lambda}H_{u}\cdot H_{d}S + \frac{1}{3}\kappa
A_{\kappa}S^{3}+\mathrm{h.c.}\big) ,
\end{eqnarray}
where $H_d^T\equiv(H_d^0,H_d^-)$, $ H_u^T\equiv(H_u^+,H_u^0)$ and
$g^2 = \L g_2^2+g_1^2\R/2$. Here we will not discuss the CP violation
aspects of the electroweak baryogenesis and assume $\lambda,
A_{\lambda},\kappa,A_{\kappa}\in \mathbb{R} $ for simplicity. An SU(2)$\times$U(1)
gauge is chosen such that at the physical vacuum
\begin{equation}\label{}
  v^+\equiv\braket{H_u^+}=0,\quad  v_{u}\equiv\braket{H_{u}^0}\in \mathbb{R^+}.
\end{equation}
$v^-\equiv \braket{H_d^-}=0$ is a local minimum provided that the charged
Higgs bosons have the positive mass square. Moreover,\ $\lambda
A_{\lambda}>0$ and $\lambda\kappa>0$ are assumed to realize $v_d,v_S
\in \mathbb{R^+}$. The angle $\beta$ is defined as in the MSSM:
\begin{equation}
v_d = v\cos\beta,\quad v_u = v\sin\beta,
\end{equation}
with $v \simeq 174$~GeV.


For the Higgs mass square structure, we decompose the Higgs fields as follows~\cite{Miller:2003ay}.
\begin{align}\label{GSB:basis}
H_u^0=&v_u+\f{1}{\sqrt{2}}(S_1\cos\beta+S_2\sin\beta)+\f{i}{\sqrt{2}}(P_1\cos\beta+G^0\sin\beta),\cr
H_d^0=&v_d+\f{1}{\sqrt{2}}(-S_1\sin\beta+S_2\cos\beta)+\f{i}{\sqrt{2}}(P_1\sin\beta-G^0\cos\beta),\cr
S=&v_s+\f{S_3+iP_2}{\sqrt{2}},
\end{align}
where $G^0$ is the Goldstone boson. In this basis, the doublet block has already been approximately diagonalized and $S_2$ is the SM-like which carries electroweak VEV among the doublets. In the basis $(S_1,S_2,S_3)$, the elements of the CP even Higgs mass squared matrix elements $(M_S^2)_{ij}$ are given by
\begin{align}\label{Higgs:even}
&(M_S^2)_{11}=M_A^2+(m_Z^2-\ld^2v^2)\sin^22\beta,\cr
& (M_S^2)_{12}=-\f{1}{2}(m_Z^2-\ld^2v^2)\sin4\beta,\cr
&(M_S^2)_{13}=-(M_A^2\sin2\beta+2\ld\kappa v_s^2)\cos2\beta\f{v}{v_s},\cr
&(M_S^2)_{22}=m_Z^2\cos^22\beta+\ld^2v^2\sin^22\beta,\cr
&(M_S^2)_{23}=\f{1}{2}(-M_A^2\sin^22\beta+4\ld^2v_s^2-2\ld\kappa v_s^2\sin2\beta)\f{v}{v_s},\cr
&(M_S^2)_{33}=\f{1}{4}M_A^2\sin^22\beta\L\f{v}{v_s}\R^2+4\kappa^2v_s^2+\kappa A_\kappa v_s-\f{1}{2}\ld\kappa v^2\sin2\beta.
\end{align}
where $M_A^2=2\ld v_s(A_\ld+\kappa v_s)/\sin2\beta$ defines the largest scale among these elements and is the heavy CP odd Higgs mass. We can introduce an auxiliary parameter $C_A \equiv 1-A_\ld \sin2\beta/2\mu- \kappa\sin2\beta/\ld$ to measure the mixing between singlet and doublet, i.e., $M^2_{23}=2C_A \lambda \mu v$ (The mixing $(M_S^2)_{12}$ can be safely neglected for moderate large tan $\beta$). The other light Higgs diagonal mass square $(M_S^2)_{33}$ can be written as
\bea
\label{M33}
 (M_S^2)_{33}&=&\f{1}{4}M_A^2\sin^22\beta\L\f{v}{v_s}\R^2 +4\kappa^2v_s^2 \L 1+\f{1}{R_\kappa} \R- \f{1}{2}\ld\kappa v^2\sin2\beta, \nonumber \\
 &=& - \f{1}{2}(M_S^2)_{23} \L \f{v}{v_s} \R+4\kappa^2v_s^2 \L 1+\f{1}{R_\kappa} \R+\lambda^2 v^2- \ld\kappa v^2\sin2\beta
 \eea
and we will use this formula again and again in later discussions. Here $R_\kappa \equiv 4 \kappa v_s / A_\kappa$ is a dimensionless critical variable defined for SFOEWPT.  

For the CP odd Higgs boson, $A_\kappa$ is theoretically upper bounded in order to keep the CP odd singlet-like Higgs mass squared $({\cal M}_{P}^2)_{22}$ positive~\cite{nmssm_review}:
\begin{eqnarray}
({\cal M}_{P}^2)_{11} & = & M^2_A \; , \nonumber\\
({\cal M}_{P}^2)_{22} & = & \f{1}{4} M_A^2 \sin^2 2 \beta \left( \f{v}{v_s} \right)^2 - \f {3}{2}\lambda \kappa v^2 \sin 2 \beta - \f {12 \kappa^2 v_s^2}{R_\kappa} \; , \nonumber\\
({\cal M}_{P}^2)_{12} & = & \f{1}{2} M_A^2 \sin 2 \beta \left( \f{v}{v_s} \right)  \ , 
\label{oddmassmatrix}
\end{eqnarray}
where $({\cal M}_{P}^2)_{11}$ corresponds to the mass squared $M_A^2$ of the only CP odd Higgs in the MSSM.



\subsection{Effective potential at Finite Temperature}

The starting point for the perturbative analysis of EWPT is the
finite temperature effective potential. Up to one-loop order, it
takes the form of
\begin{equation}\label{effpotential}
V(\varphi_l,T) = V_0(\varphi_l) + V_1(\varphi_l,T) + V_{daisy}(\varphi_l,T).
\end{equation}
where $\varphi_l$, $l=d,\,u,\,s$ are the classical field variables
corresponding to $H_d^0,\,H_{u}^0,$ and $S$. The tree-level part
$V_0$ follows directly from the Higgs potential in
Eq.~(\ref{treeHiggs}). We realize that our analysis at this precision may be subject to corrections from high order and the issue of gauge dependence \cite{Patel:2011th, Wainwright:2012zn} and a more complete analysis is left to a future study.

The one loop part $V_1$ consists of the Coleman-Weinberg potential
at zero temperature and thermal corrections at finite
temperature~\cite{Quiros}:
\begin{equation}\label{Vc1}
 V_{1} = \sum_i \frac{(-)^{2s_i}n_i}{64\pi^2}m_i^4(\varphi_l)\left(\ln\frac{m_i^2(\varphi_l)}{Q^2}
 -\frac{3}{2}\right) + \frac{T^4}{2\pi^2} \sum_{i} (-)^{2s_i}n_i J_i \left(\frac{m_i^2}{T^2}\right)
\end{equation}
where $i$ runs over all particles in the model, with each having degrees
of freedom $n_i$, field-dependent mass $m_i(\varphi_l)$ and spin
$s_i$. $J_i$ is the thermal integral function $J_{B(F)}$ for bosons
(fermions)
\begin{equation}
 J_{B,F}(y^2) = \int_0^\infty dx\;x^2 \ln(1\mp e^{-\sqrt{x^2+y^2}}).
\end{equation}
It tends to be zero in the non-relativistic limit, i.e., $y^2\gg1$. By contrast, it has the high temperature expansion and in particular gives rise to the well known thermal cubic term in Eq.~(\ref{Vc1}), given that $i$ is a boson. Here we work in the Landau gauge and in the $\overline{DR}$ scheme.  As for $V_{daisy}$, it is the daisy resummation
contributions from the longitudinal components of gauge bosons and
the scalar bosons~\cite{Parwani:1991gq,Carrington:1991hz,Arnold:1992rz}
\begin{equation}
 V_{daisy} = -\frac{T}{12\pi}\sum_b n_b(\overline{m}_b^3(\varphi_l,T)-m_b^3(\varphi_l)),
\end{equation}
where $\overline{m}_b$ is the thermal mass.

Finally, it should be emphasized that in our analysis, we use the three VEVs $v_l$ as the inputs   and eliminate the Higgs soft masses via the minimization conditions for the three field variables $\varphi_l$. Concretely, at one-loop order, they are given by
\begin{align}\label{tad}
\nonumber &m_{H_d}^2=\lambda(A_\lambda+\kappa v_S)v_S \tan\beta -\lambda^2(v_S^2+v^2 \sin^2\beta)
-\frac{\bar{g}^2}{2}v^2 \cos2\beta -\frac{1}{2v_1}\frac{\partial V_1(T=0)}{\partial\varphi_d}\Big| _{\varphi_l=v_l},\\
\nonumber&m_{H_u}^2=\lambda(A_\lambda+\kappa v_S)v_S \cot\beta-\lambda^2(v_S^2+v^2 \cos^2\beta)
+\frac{\bar{g}^2}{2}v^2 \cos2\beta -\frac{1}{2v_2}\frac{\partial V_1(T=0)}{\partial\varphi_u}\Big| _{\varphi_l=v_l},\\
\nonumber&m_S^2=\lambda A_\lambda\frac{v^2}{2v_S}\sin2\beta -\kappa A_\kappa v_S-\lambda^2v^2-2\kappa^2
  v_S^2+\lambda\kappa v^2 \sin2\beta-\frac{1}{2v_S}\frac{\partial V_1(T=0)}{\partial\varphi_S}\Big| _{\varphi_l=v_l}.\\
\end{align}

\section{Electroweak Phase Transition in the NMSSM}

With previous preparations, in this section we study EWPT in this model. It is well known that successful EWBG requires a SFOEWPT, namely $\gamma_c\equiv v_c(T_c)/T_c\gtrsim 0.9$. For a dedicated study of this condition based on gauge invariant quantities, see Ref. \cite{Patel:2011th}. Here $T_c$ is the critical temperature of SFOEWPT, with order parameter $v_c(T_c)$. In the SM, the lattice simulation indicates that its EWPT is actually a crossover, failing to achieve any jumps in terms of order parameter. In the MSSM, in particular after the discovery of the 126 GeV Higgs boson, a single light stop alone would be ruled out by the current Higgs data because of too large enhancement on the Higgs production rate from gluon fusion. Nevertheless, a second colored light scalar can not only reduce the gluon Higgs effective operator, but also enhance the EWPT strength \cite{Huang:2012wn}. While the NMSSM, by virtue of its tree-level effects, provides a simple way to enhance $\gamma_c$. Such effects have been studied by many groups before~\cite{Pietroni:1992in,Davies:1996qn,Huber:2000mg,Menon:2004wv,Profumo:2007wc,Funakubo:2005pu, Balazs:2013cia, Carena:2011jy}, but a detailed general analysis of SFOEWPT after the Higgs discovery is still absent and we fill the gap in this paper.

We will first introduce three types of EWPT and then propose a new way to investigate $\gamma_c$ from the zero temperature Higgs effective potential. Following this way, we make an anatomy of each type, giving semi-analytical treatment on tree-level effects and qualitative analysis of loop corrections. It is found that the latter plays a robust role in SFOEWPT, despite of the dominated tree-level effects.

\subsection{Vacua energy gap and SFOEWPT}

The NMSSM contains three Higgs fields and thus possesses a rich vacua structure, which leads to a variety of EWPT patterns. There are mainly three patterns~\cite{Funakubo:2005pu,Espinosa:2011ax}, classified by the course of the phase transition from the symmetric phase $\Omega_0$ to the EW symmetry breaking phase $\Omega_{\rm EW}$ (here we denote various phases with their VEV's):
\begin{description}
\item[Type-I: \,$\Omega_0$ $\Rightarrow$ $\Omega_{\rm S}$ $\Rightarrow$ $\Omega_{\rm EW}$] At high temperature, the universe is in the symmetric phase. As the universe cools down, it may transit to the vacuum locating at the singlet subspace, i.e., $\Omega_S$. As $T$ further decreases to the critical temperature $T_c$, $\Omega_{\rm S}$ degenerates with $\Omega_{\rm EW}$ and then the universe transits into the phase $\Omega_{\rm EW}$ \footnote{Notice that if $<S> =0$ in $\Omega_{\rm EW}$ this would induce a symmetry non-restoration effect in the $S$ direction \cite{Cheung:2013dca}. }.
  
\item[Type-II:\,$\Omega_0$ $\Rightarrow$ $\Omega_{\rm U}$ $\Rightarrow$ $\Omega_{\rm EW}$] Type-II transition passes the intermediate phase $\Omega_{\rm U}$ with $H_u$ developing a VEV first.  Here only the first step are relevant for EWPT, which recovers the SM case except that $S$ contributes to the thermal cubic terms~\cite{Quiros}. However, in this case, SFOEWPT requires large interactions between Higgs and singlet which induces sizable mixing between the two. This will change the transition type into Type-III. Generally, Type-II  is hardly strong so we will not discuss this type.

  \item[Type-III:\,$\Omega_0$ $\Rightarrow$ $\Omega_{\rm EW}$]
  The EW symmetry breaking vacuum develops first, and thus the universe in the symmetry phase transits directly into the phase  $\Omega_{\rm EW}$. It is worth pointing out that even though the transition does not undergo other phases, there are still extra local minima at $T=0$. In particular, there usually exits  local minima in the singlet subspace which makes the vacua structure Type-I-like. We will turn back to this point in later discussions. 
\end{description}

\begin{figure}[htb]
\centering
\includegraphics[width=2.8in]{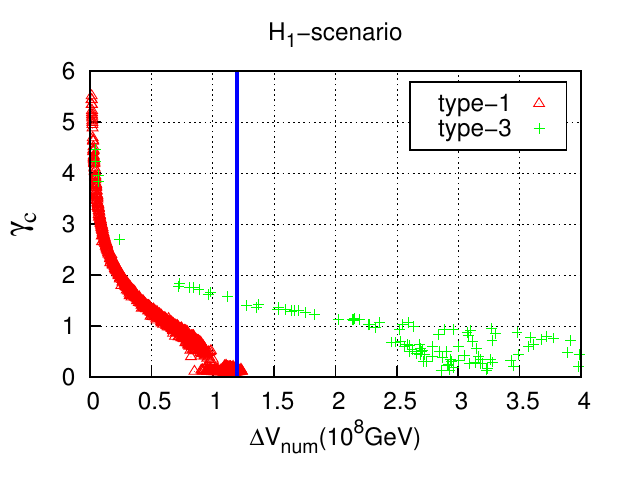}
\includegraphics[width=2.8in]{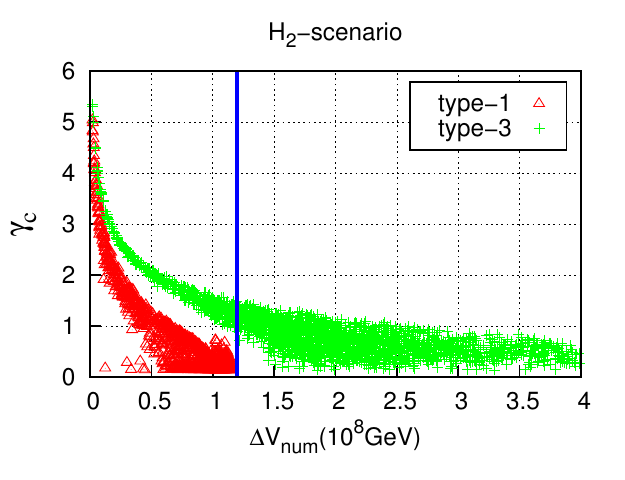}
 \caption{Strong correlation between the strength of EWPT and the vacua energy gap $\Delta V$ at $T=0$. Left panel: $H_1-$ scenario; Right panel: $H_2-$ scenario. The vertical line stands for the $\Delta V = v^2 m_h^2 /4 = 1.18 \times 10^{8}$ GeV limit. }\label{deltaV}
\end{figure}

Vacua structure at $T=0$ should encode information on EWPT. For instance, the effective potential   in Type-I is likely to have a metastable vacuum $\Omega_{\rm S}$ besides the EW vacuum $\Omega_{\rm EW}$, with vacua energy gap $\Delta V\equiv V_{\rm S}-V_{\rm EW}$. The $T$-dependent terms in the finite temperature potential need to smooth out this gap as $T$ increases, until the critical temperature. Accordingly, a smaller $\Delta V(T=0)$ may imply a lower $T_c$ thus a larger $\gamma_c$. This conjecture is confirmed by our final numerical results shown in Fig.~\ref{deltaV}. In the three-dimension field space, developing an analytical expression for $\gamma_c$ is mission impossible, except for some simplified cases like in the PQ-limit~\cite{Carena:2011jy,Balazs:2013cia}. Therefore, our observation provides an important guideline for achieving a larger $\gamma_c$.

This general correlation between the phase transition strength and the vacua energy gap $\Delta V$ can be understood in the $\Delta V \rightarrow 0$ limit. Let's consider the vacua energy gap between the symmetry phase and the broken phase:
\begin{eqnarray}
\Delta V&=& \left(V_{\rm sym}-V_{\rm EW}(v_0)\right)|_{T=0} \nonumber \\
&\simeq & V_{\rm sym}(T_{c})-V_{\rm EW}(T=0,v_0)  \nonumber \\ 
&= & V_{\rm EW}(T_{c},v_{c})-V_{\rm EW}(T=0,v_0) \nonumber \\
&\simeq  & T_{c}\frac{\partial V}{\partial T}(T=T_r,v_0)
\end{eqnarray}
where in the second line, we have used the approximation $ V_{\rm sym}(T_{c})\simeq V_{\rm sym}(T=0)$ in the $\Delta V \rightarrow 0$ limit which holds exactly if the symmetric phase is the origin of $\phi$ or is a good approximation if the symmetric vacuum is of a weakly coupled singlet. The third line comes from the degenerate vacua condition at the critical temperature. In the last line, we set $v_{c}\approx v_0$ and use the Lagrange mean value theorem and $T_r \in [0, T_c]$. Thus,
\begin{equation}
\frac{v_{c}}{T_{c}} \simeq \left( v \frac{\partial V}{\partial T}\Big|_{T=T_c} \right) \frac{1}{\Delta V}
\end{equation}

The dependence of the phase transition strength on $\Delta V$ can be revealed more explicitly in the simplified model of NMSSM in the PQ-limit~\cite{Carena:2011jy}. In this model,
\begin{equation}
\Delta V = \frac{v^4}{2}\left(\tilde{\lambda}-\frac{2\tilde{a}^2m^2_{s}}{(m^2_{s}+\lambda^2 v^2)^2}\right)
\end{equation} 
For a extremely small $\Delta V $, $v_{c}\approx v$, and the phase transition strength can be rewritten as:
\begin{equation}
\frac{v_{c}}{T_{c}}=E/\left(\tilde{\lambda}-\frac{2\tilde{a}^2m^2_{s}}{(m^2_{s}+\lambda^2 v^2\cdot \frac{v^2_{c}}{v^2})^2}\right) \simeq \frac{v^4 E}{2 \Delta V} \,
\end{equation}
where $E$ is the coefficient of thermal cubic term $\phi^3 T$.


\subsection{Anatomy of Type-I}

In this subsection we will lead the way to generate a smaller energy gap $\Delta V(T=0)$ in Type-I. A full understanding requires analysis at both tree-level and loop-level. We also give numerical results, which are consistent with those semi-analytical understandings.

\subsubsection{Tree-level analysis}\label{Tree-level analysis}

First of all, let's investigate the vacuum energy $V_{\rm {EW}}$ of ${{\Omega}_{\rm{EW}}}$. Substituting Eq.~(\ref{tad}) into Eq.~(\ref{treeHiggs}), one can eliminate the Higgs soft masses in $V_{\rm {EW}}$, and then $V_{\rm EW}$ can be divided into three parts, $V_{\rm {EW}}^H$, $V_{\rm {EW}}^S$ and $V_{\rm EW}^{HS}$. The first part denotes the contribution completely from the Higgs doublets
\begin{equation}\label{VS}
V_{\rm {EW}}^H=-\f{v^2}{4}\L g^2v^2\cos^22\beta+\ld^2 v^2\sin^22\beta \R\simeq -\f{v^2m_h^2}{4} .
\end{equation}
To derive the second approximation we have used nothing but Eq.~(\ref{mh}) where we neglect the mixing effects for the Higgs boson mass. As one can see, $V_{\rm {EW}}^H$ is definitely negative. Moreover, its value is determined by the Higgs quartic coupling thus related to the SM-like Higgs boson mass, $m_h\simeq126$ GeV. Therefore, this part is almost fixed to be around $-1.18\times 10^8\,\rm {GeV}^4$. The second part $V_{\rm {EW}}^S$ is the contribution from the singlet sector, taking the form of
\begin{equation}\label{VES}
V_{\rm {EW}}^S=-\f{1}{3}{\kappa A_\kappa}v_s^3-\kappa^2 v_s^4.
\end{equation}
The third part $V_{\rm EW}^{HS}$ is a result of the doublet-singlet mixing, and it can be casted into a simple form,
\begin{equation}\label{}
V_{\rm EW}^{HS}=-C_A\mu^2v^2.
\end{equation}
As we have mentioned before, $C_A$ measures the mixing between singlet and doublet $M_{23}^2 = 2 C_A \lambda \mu v$. From the current Higgs data, we expect this auxiliary parameter $C_A$ is usually much smaller than 1~\cite{Kang:2013rj}~\footnote{The energy $V_{\rm EW}^{HS}$ is proportional to $C_A$ while accompanied by $\mu^2$, so in the large $\mu$ region this term may have some influence.}, so the singlet or the SM-like Higgs mass is not largely pushed down in the $H_2-$ or the $H_1-$scenario respectively. This fact will help us to simplify discussions and furthermore find out a crucial variable $R_\kappa$ which has a close relation with the vacua energy gap $\Delta V$ and the EWPT strength $\gamma_c$.

Next we discuss $V_{\rm S}$, the tree-level potential energy of the absolute minimum $u_s$ in the singlet subspace \footnote{$u_s$ is the zero temperature correspondence of $\Omega_{\rm S}$, and it is not necessary to be a metastable vacuum. We also refer to it as $\Omega_{\rm S}$ in the following.}. To compare with $V_{\rm EW}$, it is convenient to eliminate $m_S^2$ in favor of $v_S$ through the third equation of Eq.~(\ref{tad}), rewritten as
\begin{align}\label{ms2app}
m_S^2=-C_A\,\ld^2 v^2 -\kappa A_\kappa v_s-2\kappa^2 v_s^2.
\end{align}
Then from the potential with only singlet $S$
\begin{align}\label{VS}
V(S)=& m_S^2 S^2 + \frac{2}{3} \kappa A_\kappa S^3 + \kappa^2 S^4 ,
\end{align}
we can get
\begin{align}\label{V(S)}
V_{\rm S}=&\left[-\kappa A_\kappa\L v_s-2 u_s/3\R - \kappa^2\L 2v_s^2-u_s^2\R \right]u_s^2-C_A\ld^2u_s^2 v^2.
\end{align}
It is also illustrative to express $V_{\rm S}$ in terms of the inputs only,
\begin{align}\label{Vs}
V_{\rm S}=-\f{A_\kappa^4}{384 \kappa^2}\L 1 + \sqrt{1 - 8 x_\kappa}\R^2 \L 1+
     \sqrt{1 - 8 x_\kappa} - 12 x_\kappa\R,
\end{align}
which holds for $x_\kappa\equiv m_S^2/A_\kappa^2<1/8$, see more details in Appendix.~\ref{appa-min}. $V_{\rm S}$ is definitely negative for $x_\kappa<1/9$. Moreover, it is an even function of both $A_\kappa$ and $\kappa$. For a given $x_\kappa$, Eq.~(\ref{Vs}) indicates that $V_{\rm S}$ becomes more negative as $A_\kappa^2/\kappa$ (or $A_\kappa/\kappa$ to some extent) increases.

With all the above expressions of potential energy, we proceed to discuss the vacua energy gap at tree level, which is found to be related to the deviation of $u_s$ from $v_s$. To see it, consider the small deviation case and write $u_s=(1+\delta)v_s$ ($|\delta|\ll 1$), then the gap is approximated as
\begin{align}\label{V:app}
\Delta V_{\rm tree}=& V_{\rm S}-V_{\rm {EW}}^S-V_{\rm {EW}}^{HS} - V_{\rm EW}^H \notag \\
 \simeq & \frac{v^{2}m_{h}^{2}}{4}-C_{A}\lambda^{2}v^{2}(u_{s}^{2}-v_{s}^{2})+\kappa^{2}(v_{s}^{2}-u_{s}^{2})^{2} \nonumber \\
 + & \frac{1}{3}\kappa A_{\kappa}\left[2u_{s}^{2}(u_{s}-v_{s})+v_{s}(v_{s}^{2}-u_{s}^{2})\right] \\
\approx & {4}\delta^2\kappa^2 \L 1+1/R_\kappa \R v_s^4-2\delta\, C_A v^2\mu^2 +\f{v^2m_h^2}{4},
\end{align}
Obviously,  $\Delta V_{\rm tree}$ goes to the doublet limit ${v^2m_h^2}/{4}$ as $\delta\ra 0$. In other words, a substantial deviation is necessary to decrease the energy gap away from the doublet limit. In fact, $\Delta V_{\rm tree}$ can be even negative (we will see this soon later). A negative $\Delta V_{\rm tree}$ is somewhat welcome since loop correction will be found to favor uplifting $V_{\rm S}$ relative to $V_{\rm EW}$.

One could have a closer inspection into the deviation. Consider the minima structure of the singlet subspace at tree level, whose details are listed in Appendix.~\ref{appa-min}, its absolute minimum locates at the origin or
\begin{align}\label{US}
u_s=\f{- A_\kappa}{4\kappa}\L 1+\sqrt{1-8x_\kappa}\R,
\end{align}
In Type-I the latter is just the case, which requires $x_\kappa<1/9$. Using Eq.~(\ref{ms2app}) one can rewrite $x_\kappa$ as
\begin{align}\label{xkappa}
x_\kappa=\f{1}{8}-\f{1}{8}\L1+R_\kappa\R^2-C_A \ld^2v^2/A_\kappa^2.
\end{align}
In the limit $C_A\ra0$, one gets the following simple relation between $u_s$ and $v_s$:
\begin{align}\label{usvs}
u_{s} & \approx
\begin{cases}
-v_{s}(1+2/R_{\kappa})+{\cal O}(C_{A}) & \text{if $R_{\kappa}\gtrsim-1$}; \\
v_{s}+{\cal O}(C_{A}) & \text{if $R_{\kappa}\lesssim-1$}.
\end{cases}
\end{align}
which shows that $u_s$ usually deviates from $v_s$ significantly in the first case while in the second case they should be close to each other, given suppressed corrections from nonzero $C_A$. Note that at the leading order ${\cal O}(C_A)$ is given by $-C_A{\ld^2 v^2}/{|1+R_\kappa|\kappa A_\kappa}$, which indicates that the approximation breaks down for $R_\kappa$ near $-1$. In this case,  a positive $C_A$ in $x_{\kappa}$  can also generate a deviation. 

Arguably, a substantial VEV deviation, i.e., for the first case in Eq.~(\ref{usvs}), tends to drive $\Delta V_{\rm tree}<0$. Notice that in the decoupling limit $C_A\ra0$, $S=v_s$ is always either a minimum or maximum ($A_\kappa < 0$ and $R_\kappa\gtrsim-1$) in the singlet subspace since the first derivative of $V(S)$ from Eq.~(\ref{V(S)}) over $S$ is zero \footnote{Turning on a small $C_A$ will make $\Omega_{\rm EW}$ stable and shift $S=v_s$ away from being a maximum.}. In the latter case, it is not surprising that $V_{\rm EW}^S(v_s)>V_{\rm S}(u_s)$; in the former case, a large negative $V_{\rm S}(u_s)$ is also possible for $u_s\lesssim -v_s$ ($R_{\kappa}> 0$). Remind that the singlet-doublet mixing term is suppressed by small $C_A$, thus the above difference tends to dominate in $\Delta V_{\rm tree}$, rendering it negative. This is particularly true in the $A_\kappa<0$ region when $V_{\rm S}(v_s)$ is a maximum, where $R_\kappa>-1$ requires $-A_\kappa/\kappa>4v_s\gtrsim {\cal O}(\rm TeV)$ or even order of magnitude larger for a larger $\mu$. That large  $-A_\kappa/\kappa$, in terms of the naive argument below Eq.~(\ref{Vs}), renders $\Omega_S$ well below $\Omega_{\rm EW}$. Therefore substantial loop corrections are indispensable to flip the order.

\begin{figure}[!htb]
\centering
\includegraphics[width=2.8in]{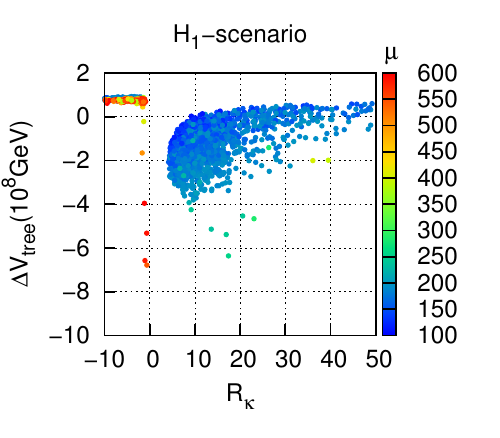}
\includegraphics[width=2.8in]{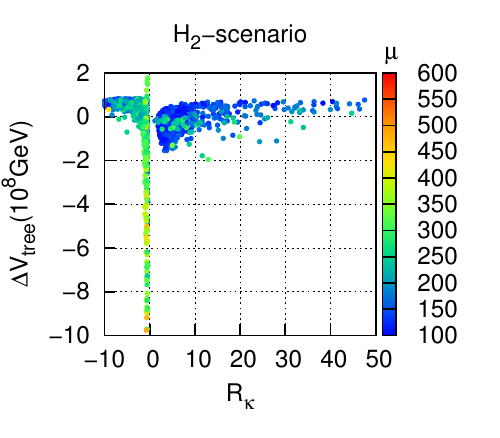}
\includegraphics[width=2.8in]{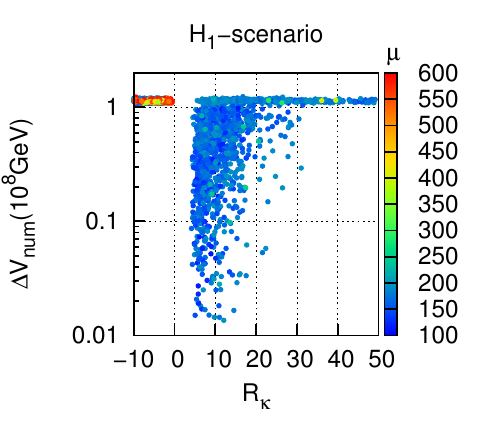}
\includegraphics[width=2.8in]{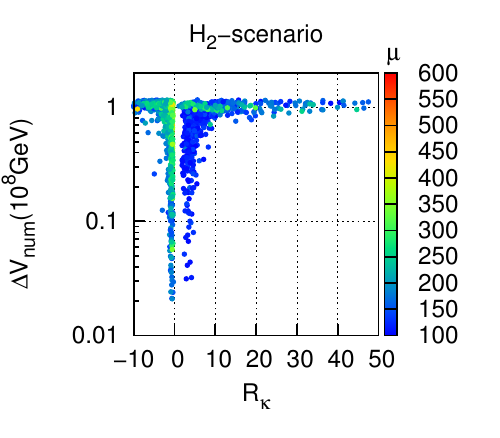}
\centering
   \caption{$\Delta V_{\rm tree}$ and $\Delta V_{\rm num}$ (with loop corrections) versus $R_\kappa$. The left(right) two figures are the plots for the $H_1$($H_2$)$-$scenario. Clearly, the large negative energy gap at the tree-level $\Delta V_{\rm tree}$ is driven back to small positive value $\Delta V_{\rm num}$ through loop corrections.}
   \label{gap_V}
\end{figure}

Before heading towards the loop corrections, let's make some observations of the tree-level results on the $R_\kappa-\Delta V_{\rm tree}$ plane (see the upper panel of Fig.~\ref{gap_V}).
They are in accord with the analysis above: (I) In $R_\kappa\ll-1$ region, $u_s \simeq v_s$, so $\Delta V_{\rm tree}$ clearly takes the doublet limit ${v^2m_h^2}/{4}$; (II) In $R_\kappa\sim-1$ region, the magnitude of $\Delta V_{\rm tree}$ can blow up, in particular within the window $-1<R_\kappa<0$ and for a relatively large $\mu$; (III) In $R_\kappa>0$ region, as argued before, $\Delta V_{\rm tree}$ can also be negative and of order of a few $10^{8}\rm\,GeV^4$, significantly smaller than the case (II). More complementary analysis is left to the part of numerical study.   

\subsubsection{Loop-level analysis}\label{loopana}

Previously it is shown  at tree level $\Omega_{\rm S}$ usually lies above $\Omega_{\rm EW}$. Here we will demonstrate that the tree-level order is going to be flipped by loop effects, which tend to lower $\Omega_{\rm EW}$ but lift up $\Omega_S$. In the $\overline{\rm DR}$ scheme, the former is mainly ascribed to the remnant of Coleman-Weinberg potential after correcting the Higgs soft mass terms, while the latter is mainly due to the shift of $u_s$. In the following we describe their details respectively.


On one hand, loop effects can lower $\Omega_{\rm EW}$. To offset the shift of VEV's in $\Omega_{\rm EW}$ due to  Eq.~(\ref{Vc1}), we may need to add the corresponding Higgs quadratic term, $-\f{1}{2}\Delta m_l^2 \varphi_l^2$, with $\Delta m_l^2$ determined to be
\begin{align}\label{shift}
\Delta m_l^2=2\sum_i A_i
 \f{m_i^2(v_l)}{v_l}\left[m_i^2(\varphi_l)\right]'_{v_l}
\left(L_i(v_l)
 -1\right).
\end{align}
We have introduced $A_i\equiv (-)^{2s_i}n_i/64\pi^2$ and $L_i\equiv \ln\frac{m_i(v_l)^2}{Q^2}$ for short. As a consequence, the remnant of Coleman-Weinberg potential, $V_{\rm CW}-\f{1}{2}\Delta m_l^2 \varphi_l^2$, results in a shift to the tree-level $V_{\rm EW}$ vacuum energy:
\begin{align}\label{}
V_{\rm CW}^{R}\equiv\sum_i A_im_i^4(v_l) \left[L_i(v_l)
-\f{v_l}{m_i^2(v_l)}\left[m_i^2(\varphi_l)\right]'_{v_l}
\left(L_i(v_l)
 -1\right)\right].
\end{align}
The above expression can be simplified greatly for two limits of $m_i^2(v_l)$:
\begin{description}
  \item[Strong $v_l-$dependence] In this limit the mass of particle $i$ dominantly originates in coupling to Higgs fields, such as the SM-particles and Higgsinoes. Then we have
      \begin{align}\label{strong}
V_{\rm CW}^R(v_l)=-\sum_i A_im_i^4(v_l)
\left(L_i(v_l)-\f{1}{2}\right).
\end{align}
Bare in mind that we have fixed $Q=2$ TeV, thus the relatively light fermions, e.g., top quark and Higgsinos, contributes a positive $V_{\rm CW}^R$ and make for flipping. The resulted decrease in $V_{\rm EW}$ can be up to order $10^9\,\rm GeV^4$ for a heavy $\mu\sim500$ GeV. By contrast, the light weak gauge bosons hamper for flipping but numerically it is unimportant for their lightness.

  \item[Weak $v_l-$dependence] Some particles like stop have large (soft) mass terms, so they typically have quite weak dependence on $v_l$ \footnote{In our paper we decouple the stop in the thermal plasma but include their radiative corrections to the Higgs potential.}. In this limit one may write $m_i^2(v_l)=m_i^2\L 1+ f(x)\R$ with $x=v_l/m_i$ and $f(x)\ll1$. With that, we get an approximation
       \begin{align}\label{weak}
V_{\rm CW}^R\approx& A_im_i^4 \L L_i-3/2\R-\cr
&
A_im_i^4\left[\L1-L_i\R \left( 2f(x)-xf'(x)\right)-L_if(x)\L f(x)-xf'(x)\R\right],
\end{align}
Here $L_i \equiv \ln \frac{m_i^2}{Q^2}$. The term in the first line is a constant thus contributing null to the energy gap. While for the second line, heavier CP even/odd Higgs bosons with their mass dependences on the VEVs ($f(x) \sim x$) will benefit the reduction of $V_{CW}^R$ at the VEVs. If $f(x)=\ld_l^2 x^2$ (stop without trilinear soft mixing), the leading $x-$dependence in the second row will vanish, with energy shift proportional only to powers of $v_l$, i.e., $A_i  L_i \ld_l^4v_l^4$. 
\end{description}
In summary, viewing from our particle spectrum, loop effects tend to decrease energy of $\Omega_{\rm EW}$. In the following we discuss the $u_s-$shift effect on $V_{\rm S}$.

On the other hand, loop effects can lift $V_{\rm S}$ up. Here the discussion is different from the previous case, because in $\Omega_S$ the singlet VEV changes after loop corrections and the effective radiative potential plays an important role. It induces a shift to $m_S^2$, inherited from the previous discussions in $\Omega_{\rm EW}$. On top of that, it affects other tree-level couplings, as can be seen by expanding $V_{\rm CW}(S)$ into polynomials of $S$. We give corresponding typical examples: The heavy Higgs bosons (still lighter than $\mu$) and Higgsinos with mass $\ld s$ increase $m_S^2$ and $\kappa$ by an amount, respectively, $\sim {\rm TeV}^2/16\pi^2$ and
\begin{align}\label{shiftkappa}
 \kappa\ra \kappa \L 1+(1-L)\f{{\ld^4}}{8\pi^2\kappa^2}\R^{1/2}.
\end{align}
From Eq.~(\ref{US}) and Eq.~(\ref{Vs}) we know that both $|u_s|$ and $|V_{\rm S}|$ are monotonically decreasing functions of $x_\kappa$ (and $\kappa$ as well from Eq.~(\ref{xkappa}) in $C_A \rightarrow 0$ limit.). Thus, in the region with relatively small $\kappa^2$ ($\lesssim0.01$), $|u_s|$ may be decreased and negative $V_{\rm S}$ is increasing so $\Omega_S$ is lifted up. 

The loop-level numerical results on the $R_\kappa-\Delta V_{\rm tree}$ plane are shown in the lower panel  of Fig.\ref{gap_V}. The second role of lifting $V_{\rm S}$ which effectively increases $\kappa$ is crucial to flip vacua order with an especially large tree-level gap, which is characterized by $-1<R_\kappa<0$. 
As mentioned before, it is usually accompanied with a relatively smaller $\kappa$ and large $\mu$, which yields a sizable increase of $\kappa$ from Eq. (\ref{shiftkappa}). Recall that $R_\kappa\propto \kappa$, increasing $\kappa$ may drag $R_\kappa$ out the window $-1<R_\kappa<0$ and make $R_\kappa<-1$. Therefore from Eq.~(\ref{V:app}), It is reasonable for us to draw a conclusion, i.e., the loop-level gap goes to the doublet limit $m_h^2 v^2/4>0$ for those points. In this way, the tree-level order is flipped. 



\subsubsection{Numerical results}
The SFOEWPT is the result of complicated interplay among quite a few parameters, including $\ld$, $\kappa$, $A_\kappa$ etc. We have turned to the numerical methods for a global understanding, and for a cross-check with previous qualitative analysis. Using the NMSSMTOOLS package~  \cite{Ellwanger:2004xm,Ellwanger:2005dv,Ellwanger:2006rn}, we scan the parameter space of the model with constraints from various relevant experiments, including the constraints on Higgs signatures. The parameter setting is listed as follows:
\begin{align}\label{para}
 \kappa:&\,(0.01,\,0.5),\quad  \ld:\, (0.3,\,0.8),\quad  \tan\beta:\, (1.5,\,10),\cr
 A_\ld:&\,(200,\,2000)\,\rm GeV,\quad  A_\kappa:\,(-1000,\, 1000)\,\rm GeV,\quad  \mu:\, \,(100,\,600)\,\rm GeV.
\end{align}
To minimize the uncertainty from the soft spectrum and explore the EWPT properties from the genuine Higgs-singlet sector, we assume that they are irrelevantly heavy. In particular, the parameters in the stop sector are taken to be $m_{\widetilde Q}^2=m_{\widetilde t_R}^2=2$ TeV and $A_t=0$~\footnote{In fact, decoupling the stop is not merely requiring heaviness, and we additionally require a smaller $A_t$. This can be seen from Eq.~(\ref{weak}), the presence of sizable $A_t$ would make heavy stops leave appreciable effects on energy gap at loop level.}. The numerical points used in the previous sections are actually from this parameter space. We calculate the phase transition strength following the textbook way: Search for the minima of the complete one loop Higgs potential (\ref{effpotential}) at each temperature, and then find out $T_c$ and $\varphi_c$ by the degenerate vacua condition. In the following we show the parameter distributions favored by SFOEWPT, and try to give interpretations of them.

\begin{figure}[!ht]
\centering
\includegraphics[width=2.8in]{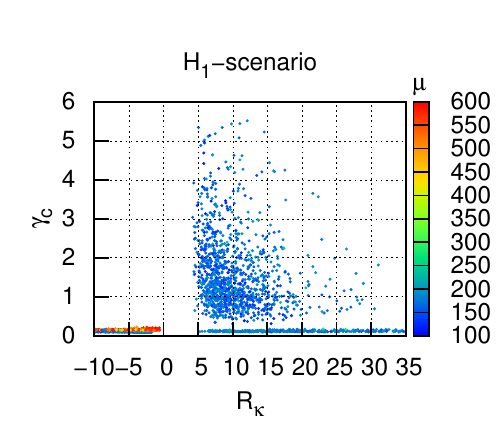}
\includegraphics[width=2.8in]{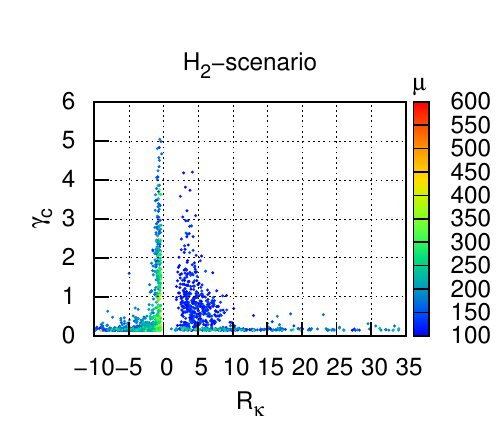}
 \caption{$R_{\kappa}$ versus $\gamma_{c}$ in Type-I transition, with color code denoting $\mu$. Left panel: $H_1-$scenario; Right panel: $H_2-$sceanrio. }\label{T1_Rkap_PTS_mu}
\end{figure}

\begin{figure}[h]
\includegraphics[width=2.8in]{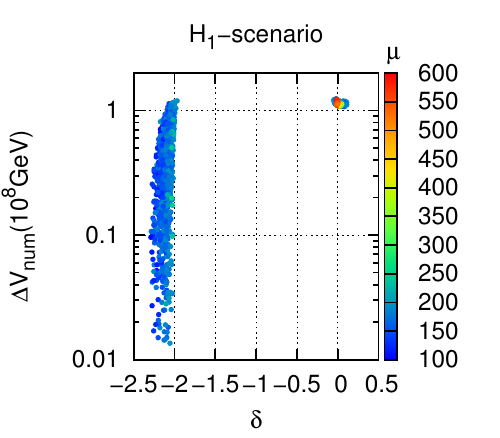}
\includegraphics[width=2.8in]{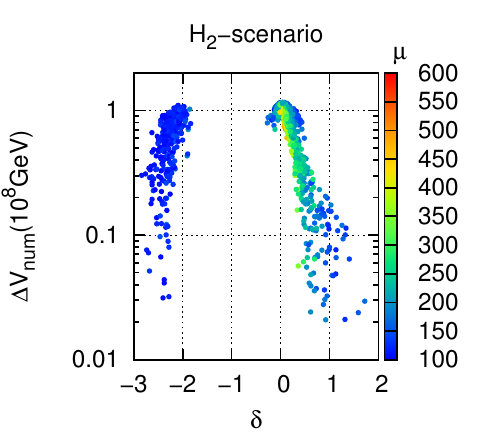}
   \caption{Loop-level gap $\Delta V_{\rm num}$ versus $\delta \equiv (u_s-v_s)/v_s$ for the Type-I transition for the $H_1-$ and $H_2-$scenario respectively. }
   \label{gap_delta}
\end{figure}

\begin{figure}[!htb]
\centering
\includegraphics[width=2.8in]{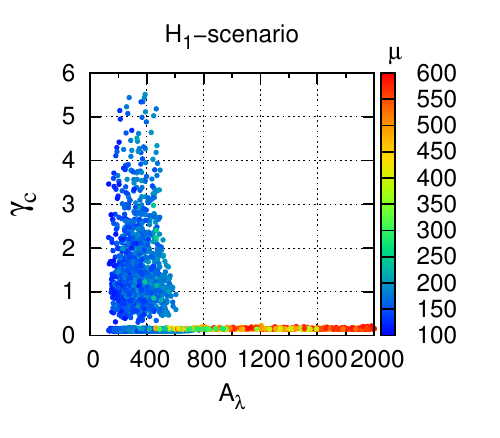}
\includegraphics[width=2.8in]{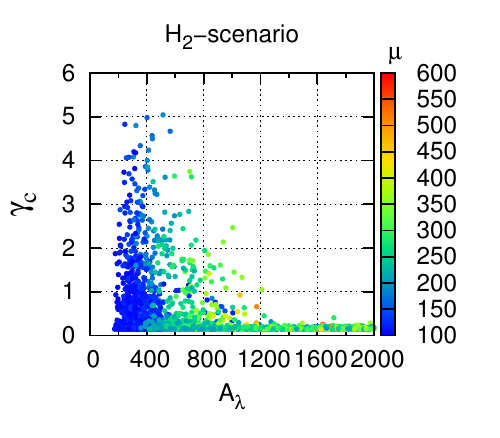}
 \caption{As in Fig. \ref{T1_Rkap_PTS_mu}, plots on the $A_{\lambda}-\gamma_{c}$ plane.}\label{T1_Alam_PTS}
\end{figure}

Firstly, we present plots on the $R_\kappa-\gamma_c$ plane in Fig.~\ref{T1_Rkap_PTS_mu}. From them one can see that $R_\kappa$ is an helpful variable to judge $\gamma_c$. In other words, the previous tree-level analysis, despite of very complicated loop corrections, still provides  valid insights. A possible large $\gamma_c$ is accommodated in two distinct regions:
\begin{description}
  \item[$R_\kappa\sim{\cal O}(1)-{\cal O}(10)$] $\gamma_c>1$ in Type-I favors the $R_\kappa>0$ (thus $A_\kappa/\kappa>0$) region, which could have a sizable $(u_s-v_s)$ deviation to avoid the doublet limit. One may wonder why large $R_\kappa$'s fail. Consider $R_\kappa\gg1$ and still take $C_A\ra0$, from Eq.~(\ref{usvs}) one gets $u_s\approx -v_s$. Then it is straightforward to derive
     \begin{align}\label{LargeR}
 \Delta V_{\rm tree}\approx m_h^2 v^2/4-16\kappa^2 v_s^4/3R_\kappa+{\cal O}(C_A).
\end{align}
So $\Delta V_{\rm tree}$, with its singlet contribution suppressed by the large $R_\kappa$, goes to the doublet limit again. Numerical results show that in $H_{2}-$scenario, $\gamma_c>1$ occurs for $R_\kappa\lesssim 10$, while in $H_{1}-$scenario $\gamma_c>1$ can still occurs for $R_\kappa$ as large as 30. 

Note that $R_\kappa$ can not be too small neither owing to phenomenological reasons. A large positive $A_\kappa$ threatens the positivity of the light CP odd Higgs boson mass (we will turn back to it later). Moreover, a very small $\kappa$ leads to a light (singlino-like) neutralino, which opens too large 126 GeV Higgs exotic decay branching ratio in the non-PQ limit. Besides, singlino-like LSP may be over-abundant because of too small annihilation cross section.  In conclusion, $A_\kappa/\kappa$ can not be too large and accordingly $R_\kappa$ gets a lower bound, about 4.0 and 2.0 in $H_{1}-$scenario and $H_{2}-$scenario respectively as shown in Fig.~\ref{T1_Rkap_PTS_mu}.

\item[$R_\kappa\sim-1$] For $R_\kappa<0$, it is not surprising that points with $\gamma_c>1$ crowd around $R_\kappa\sim -1$. On one hand, $R_\kappa\ll-1$ fails to achieve SFOEWPT owing to the doublet limit, as argued at the end of Section \ref{Tree-level analysis}. On the other hand, $R_\kappa$ cuts off before approaching to zero because $-A_\kappa$ is not allowed to be very large here owing to the singlet-like CP even Higgs boson. Therefore, $R_\kappa$ is preferred to be around -1.

$H_{1}-$scenario and $H_{2}-$scenario demonstrate a remarkable distinction in this region, namely $\gamma_c\gtrsim1$ is accommodated in the latter but not in the former. , If the point is in the $H_1-$scenario with $\gamma_c\gtrsim1$, a large $M_{33}$ term is required and from Eq. (\ref{M33}), we can see that this further requires a large $M_A$ (very large $\mu$ and $A_\lambda$, see Fig. \ref{T1_Rkap_PTS_mu} and Fig. \ref{T1_Alam_PTS}) since the term $ 4 \kappa^2 v_s^2 (1+ 1/R_\kappa)$ is close to its minimal.  

Notice that those points in $H_{1}-$scenario have obviously large $\mu$'s, which yield considerably loop corrections according to the discussion at the end of Section~\ref{loopana}. Those loop corrections would bring $\delta$ back to zero (see Fig.~\ref{gap_delta}) and therefore forbid the SFOEWPT. As a comparison, in $H_{2}-$scenario $\delta$ could have a very large deviation from zero and even approach $\delta > 1$. It is precisely those points which have a small $\Delta V$ and trigger a SFOEWPT.    
\end{description}

\begin{figure}[b]
\centering
\includegraphics[width=2.8in]{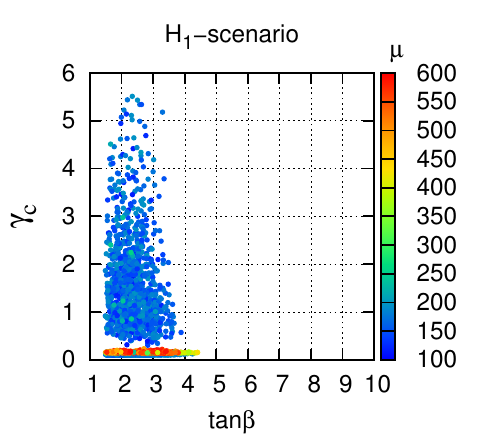}
\includegraphics[width=2.8in]{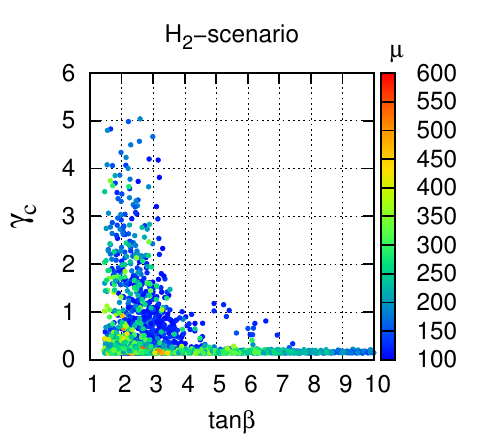}
 \caption{As in Fig. \ref{T1_Rkap_PTS_mu}, plots on the $\tan\beta-\gamma_{c}$ plane. }\label{T1_tanb_PTS}
\end{figure}
\begin{figure}[!htb]
\centering
\includegraphics[width=2.8in]{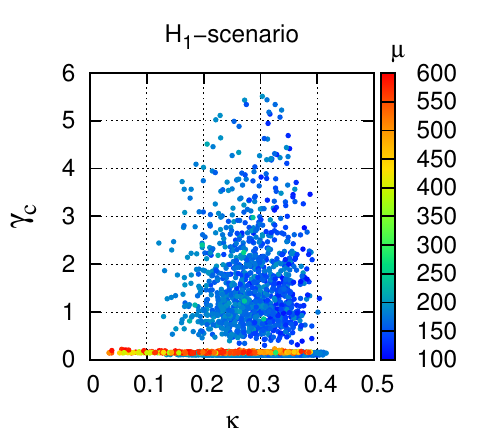}
\includegraphics[width=2.8in]{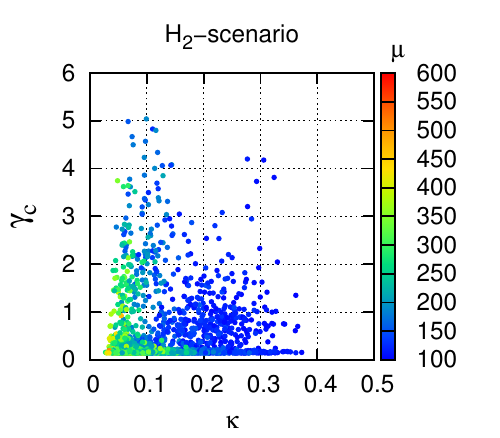}
 \caption{As in Fig. \ref{T1_Rkap_PTS_mu}, plots on the ${\kappa}-\gamma_{c}$ plane. }\label{T1_kap_PTS}
\end{figure}

\begin{figure}[h]
\centering
\includegraphics[width=2.8in]{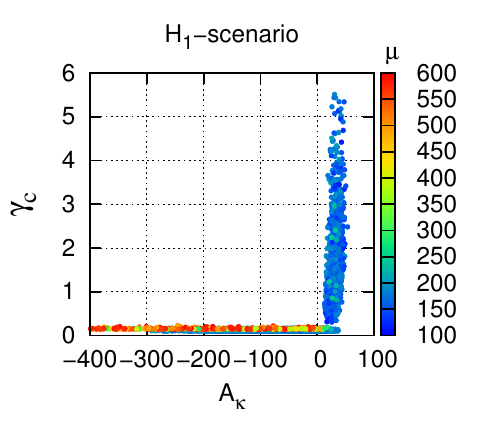}
\includegraphics[width=2.8in]{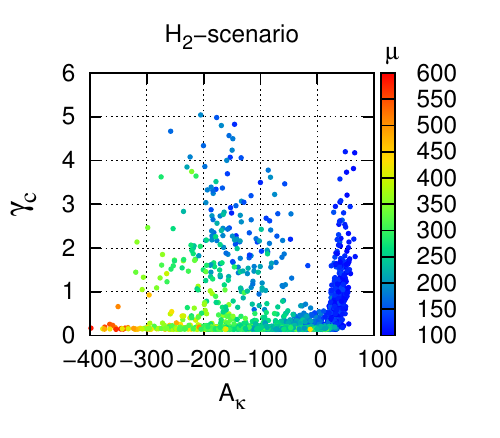}
 \caption{As in Fig. \ref{T1_Rkap_PTS_mu}, plots on the $A_{\kappa}-\gamma_{c}$ plane. }\label{T1_Akap_PTS}
\end{figure}

Next, we outline the distributions of relevant parameters favored by $\gamma_c>1$ in Type-I:
\begin{itemize}
  \item Light $\mu$ and small $\tan \beta$ are preferred (see Fig.~\ref{T1_tanb_PTS}), in particular in $H_{1}-$scenario where $\mu\lesssim250$ GeV and $\tan \beta \lesssim 3.5$. In $H_{2}-$scenario they can extend to a bit larger regions, i.e., $\mu\lesssim250$ GeV and  $\tan \beta \lesssim 5$. In this sense, Type-I agrees with the most natural NMSSM scenario~\cite{Kang:2012sy}, which have a slice of parameter space with $\ld\sim 1$, $\tan\beta\sim 1$ and $\mu\sim m_Z$. 
  \item $A_\lambda\sim(200, 600)$ GeV (see Fig.~\ref{T1_Alam_PTS}). Due to the suppressed mixing effect $|C_A| \ll 1$,  $A_\lambda$ is strongly correlated with $\mu$, i.e., $A_\lambda \approx 2\mu/\sin2\beta$.
  \item As for $\kappa$ and $A_\kappa$, most of the preferences can be traced back to the discussion on $R_\kappa= 4 \kappa v_s/A_\kappa$. We emphasize again that both $H_{1}-$ and $H_{2}-$scenario accommodate SFOEWPT for $A_\kappa>0$ , while for $A_\kappa<0$, $\gamma_c$ can hardly achieve $\mathcal{O}(1)$ in $H_{1}-$scenario (see Fig.~\ref{T1_Akap_PTS}).

\end{itemize}

\subsection{Type III: Results and analysis}

In this subsection, we turn our attention to the Type-III transition. This type of EWPT is studied in detail by the early works~\cite{Pietroni:1992in,Menon:2004wv,Funakubo:2005pu,Carena:2011jy} due to its compatibility with the near PQ symmetry limit and its one-step nature. Here we revisit this type of transition in the spirit of energy gap. Most of the analysis is similar to that of Type-I.

It is worth pointing out that Type-III arises not only in the case that the origin is indeed the absolute minimum in the singlet subspace but also in the case that the origin is metastable. The latter has a Type-I-like vacua structure but belongs to Type-III due to thermal evolution: At $T=0$, the absolute minimum in the subspace locates at $u_s\neq 0$, but as temperature increases it will exceed the origin and become energetically disfavored. So the degenerating eventually happens between the origin and $\Omega_{\rm EW}$, who determine the gap. Such type-crossing phenomenons are not difficulty to be understood. The critical case $x_\kappa=m_S^2/A_\kappa^2=1/9\ll1$ means that $m^2_{S}$ is positive and small (typically $m_{S} \lesssim 100$ GeV), and thus sensitive to temperature. Although it is unable to distinguish quantitatively this case from Type-I at $T=0$, our numerical results tell that the models of this case have a smaller gap between the origin and $\Omega_S$ than those in Type-I.

\begin{figure}[!htb]
\centering
\includegraphics[width=2.8in]{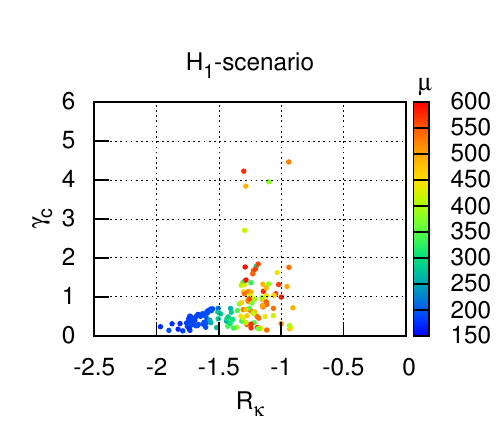}
\includegraphics[width=2.8in]{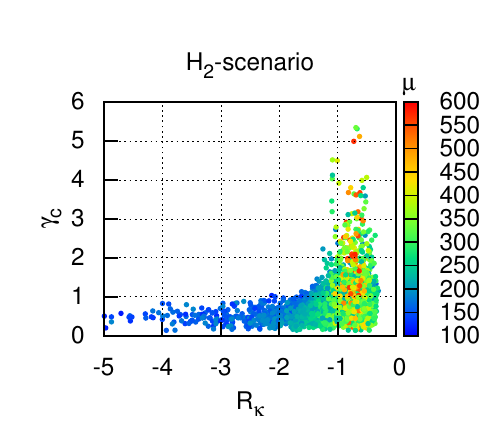}
 \caption{$R_{\kappa}$ versus $\gamma_{c}$ in Type-III transition, with color code denoting $\mu$. Left panel: $H_1-$scenario; Right panel: $H_2-$sceanrio. }\label{T3_Rkap_PTS_mu}
\end{figure}

We proceed to discuss the energy gap at tree and loop-level. This time, the tree-level gap is quite simple, given by $-V_{\rm EW}$. 
\begin{align}\label{DVtreeT3}
\Delta V_{tree} & =  -V_{\rm EW}^{H} -V_{\rm EW}^{HS} -V_{\rm EW}^{S}\nonumber \\
& \simeq  \frac{v^{2}m_{h}^{2}}{4}+C_{A}\mu^{2}v^{2}+\kappa^{2}v_{s}^{4}\left (\frac{4}{3 R_{\kappa}}+1 \right)
\end{align}
Obviously, the singlet part dominates the energy gap for a large $\mu$, and thus a small gap appeals to a large negative $A_{\kappa}$ such that $-4/3 \lesssim R_{\kappa} <0 $. For a moderate $\mu$, the mixing part becomes important and a negative $C_{A}$ can help to decrease the gap, which requires a large $A_{\lambda}$. The features outlined above are well reflected in the Fig. \ref{T3_Rkap_PTS_mu}. 

The above analysis can be well adjusted in the near PQ-limit, where the mixing part from the second term in Eq. (\ref{DVtreeT3}) could play a dominate role, so a large $A_{\ld}$ and a moderate $\mu$ play key roles in decreasing the energy gap. Notice that in Ref. \cite{Carena:2011jy} where a relatively larger $\tan \beta > 10$ region is considered, the corresponding $A_\lambda \sim (2, 5)$ TeV is even larger in order to achieve a small $\Delta V$ for SFOEWPT. We observe that Type-III can only be accommodated in a very restricted region. This is because that as soon as $R_{\kappa}$ become a bit larger, $\Omega_{\rm S}$ will become so deep that the transition changes into Type-I \footnote{the points in $R_{\kappa}\lesssim -1$ region have a negative $C_{A}$, which can relax the conflict between large $R_{\kappa}$ and Type-III, as indicated by Eq.~(\ref{ms2app}) or Eq.~(\ref{xkappa}).}. 

The loop-level analysis in Section~\ref{loopana} is also applicable to Type-III, i.e. there are mainly two kinds of loop-level effects. On the one hand, loop corrections lower $\Omega_{EW}$ and make it to be the global minimum in the same way as in Type-I. On the other hand, the loop corrections in the singlet subspace lift up $\Omega_{\rm S}$ by increasing $m^2_{S}$ and $\kappa$. However, the latter has no influence on the energy gap concerned here, and just makes $\Omega_{\rm S}$ shallow and ready to exceed the origin in thermal evolution.

\begin{figure}[!ht]
\centering
\includegraphics[width=2.8in]{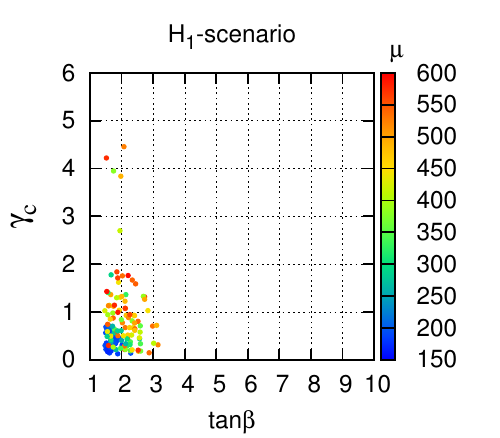}
\includegraphics[width=2.8in]{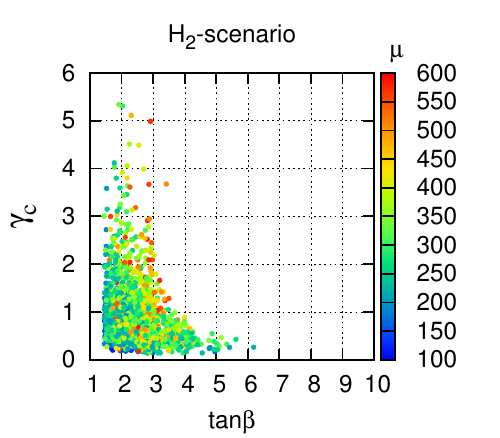}
 \caption{As in Fig. \ref{T3_Rkap_PTS_mu}, plots on the $\tan\beta-\gamma_{c}$ plane. }\label{T3_tanb_PTS}
\end{figure}

\begin{figure}[!h]
\centering
\includegraphics[width=2.8in]{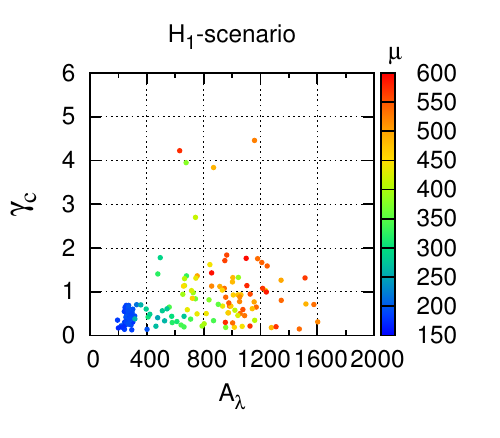}
\includegraphics[width=2.8in]{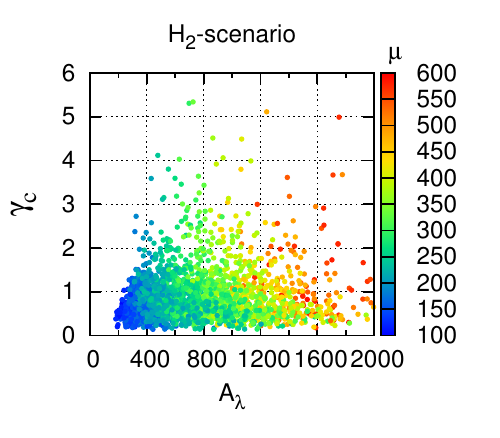}
 \caption{As in Fig. \ref{T3_Rkap_PTS_mu}, plots on the $A_{\lambda}-\gamma_{c}$ plane.}\label{T3_Alam_PTS}
\end{figure}

Finally, we summarize the parameters preference of SFOEWPT in Type-III:
 \begin{itemize}
   \item  Unlike Type-I, SFOEWPT in Type-III prefers a larger $\mu \gtrsim 250$ GeV, especially in the region $-4/3 \lesssim R_{\kappa} <0 $ (see Fig. \ref{T3_Rkap_PTS_mu}). As argued above, $\Delta V$ here is effectively decreasing by the singlet part. Similar to Type-I, $\tan \beta \lesssim 3.5$ is favored (see Fig. \ref{T3_tanb_PTS}).
   \item		The favored values of $A_{\lambda}$  here is larger than in Type-I since $A_\lambda \simeq 2\mu /\sin2\beta $ or even larger for a negative $C_{A}$(See Fig. \ref{T3_Alam_PTS}). 
   \item		Another obvious distinction between Type-I and -III can be observed in Fig.~\ref{T3_Akap_PTS}: $A_\kappa>0$ barely accommodates Type-III, not mentioning to realize $\gamma_c>1$. Eq.~(\ref{ms2app}) may provide a simple interpretation. It indicates that given a positive $A_\kappa$ we have $m_S^2\ra -\kappa A_\kappa v_s-\kappa^2v_s^2<0$ and consequently $x_\kappa<0$, which yields a large deep minimum in the singlet direction, therefore strongly favors Type-I.
   \item  The vast majority of surviving points are in $H_{2}-$scenario. As a matter of fact, seemingly $H_{1}-$scenario has a strong tension with Type-III. This may blame to its parameter configurations that is difficult to achieve a large $M_{33}$ unless $\mu$ and $A_\lambda$ are large, which forbids the SFOEWPT just like the case of Type-I.
  \end{itemize}

\begin{figure}[!ht]
\centering
\includegraphics[width=2.8in]{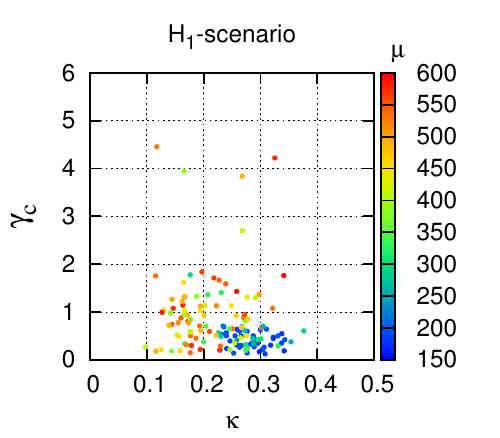}
\includegraphics[width=2.8in]{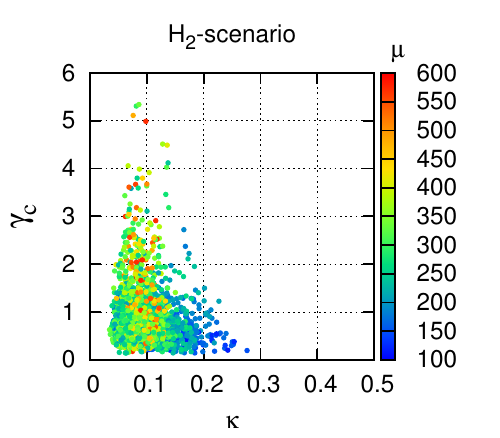}
 \caption{As in Fig. \ref{T3_Rkap_PTS_mu}, plots on the $\kappa-\gamma_{c}$ plane. }\label{T3_Akap_PTS}
\end{figure}

\begin{figure}[!ht]
\centering
\includegraphics[width=2.8in]{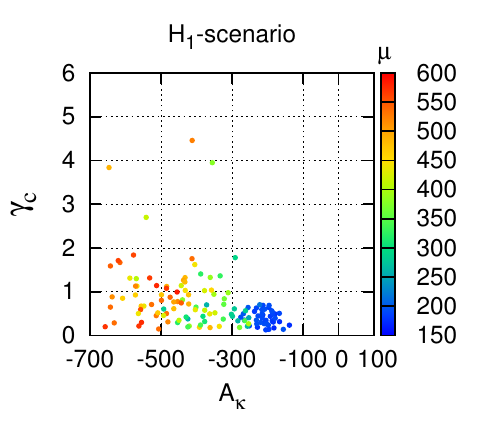}
\includegraphics[width=2.8in]{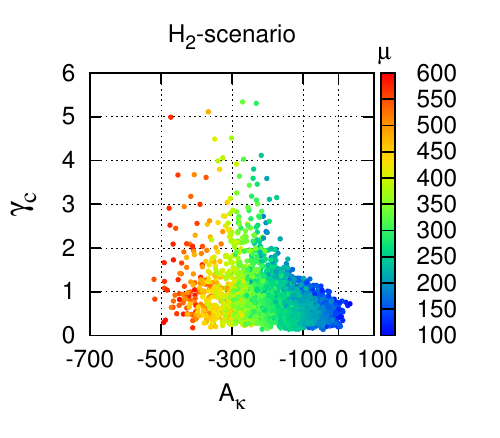}
 \caption{As in Fig. \ref{T3_Rkap_PTS_mu}, plots on the $A_{\kappa}-\gamma_{c}$ plane. }\label{T3_Akap_PTS}
\end{figure}

\section{Relevant Phenomenologies}

\subsection{Higgs Spectra}

The most remarkable connection may lie in the existence of a relatively light Higgs boson in the spectrum, though their masses are heavier than $m_{h_{SM}}/2$ otherwise the Higgs exotic decay channel would dominate and get severely constrained from the current Higgs signal rates. 

For the CP odd Higgses, it is well known that the absence of tachyon states in the Higgs CP odd sector strongly favors negative $kA_\kappa$ and now it is more constrained from the observed Higgs signal rate as mentioned above. Such a sharp contradiction substantially compresses the allowed parameter space and leads to a remarkable prediction on the Higgs spectrum.
The CP odd Higgs sector contains two physical states $a_{1,2}$. In the case of $M_A^2\gg M_Z^2$ we have
\begin{equation}
m_{a_1}^2\simeq \ld\L A_\ld+4\kappa v_s\R v_uv_d/v_s-3\kappa A_\kappa v_s
\label{ma1_mass}
\end{equation}
where the second term can be expressed as $-12 \kappa^2 v_s^2 / R_\kappa$. Thus, a positive small $R_\kappa$ drives the lighter CP odd Higgs boson mass downwards.  

For the CP even Higgses, $R_\kappa$ can also affect the singlet-dominated CP even Higgs mass, as we mentioned in Eq.~(\ref{M33}):
\bea
 (M_S^2)_{33}= 4\kappa^2v_s^2 \L 1+\f{1}{R_\kappa} \R + \cdots ,
\eea
where $\cdots$ stands for other terms either are small or not relevant here. Therefore the preference of $R_\kappa$ from SFOEWPT may result in some specific distribution in the Higgs spectrum. Especially we can see that $-1<R_\kappa<0$ is the region that the above term get its minimal and will drive down the $h_1$ mass significantly. Therefore, it is clear that in the NMSSM, the SFOEWPT will impose a specific Higgs spectra through the critical parameter $R_\kappa$:
 \begin{itemize}
   \item  $R_\kappa>0$ prefers a light CP odd Higgs mass (small $m_{a1}$) with no strong preference on $m_{h1}$.
   \item  $0> R_\kappa > -1$ prefers a light CP even Higgs mass (small $m_{h1}$ in the $H_2-$scenario) with no preference on $m_{a1}$. 
  \end{itemize}

\begin{figure}[t]
\includegraphics[width=8cm]{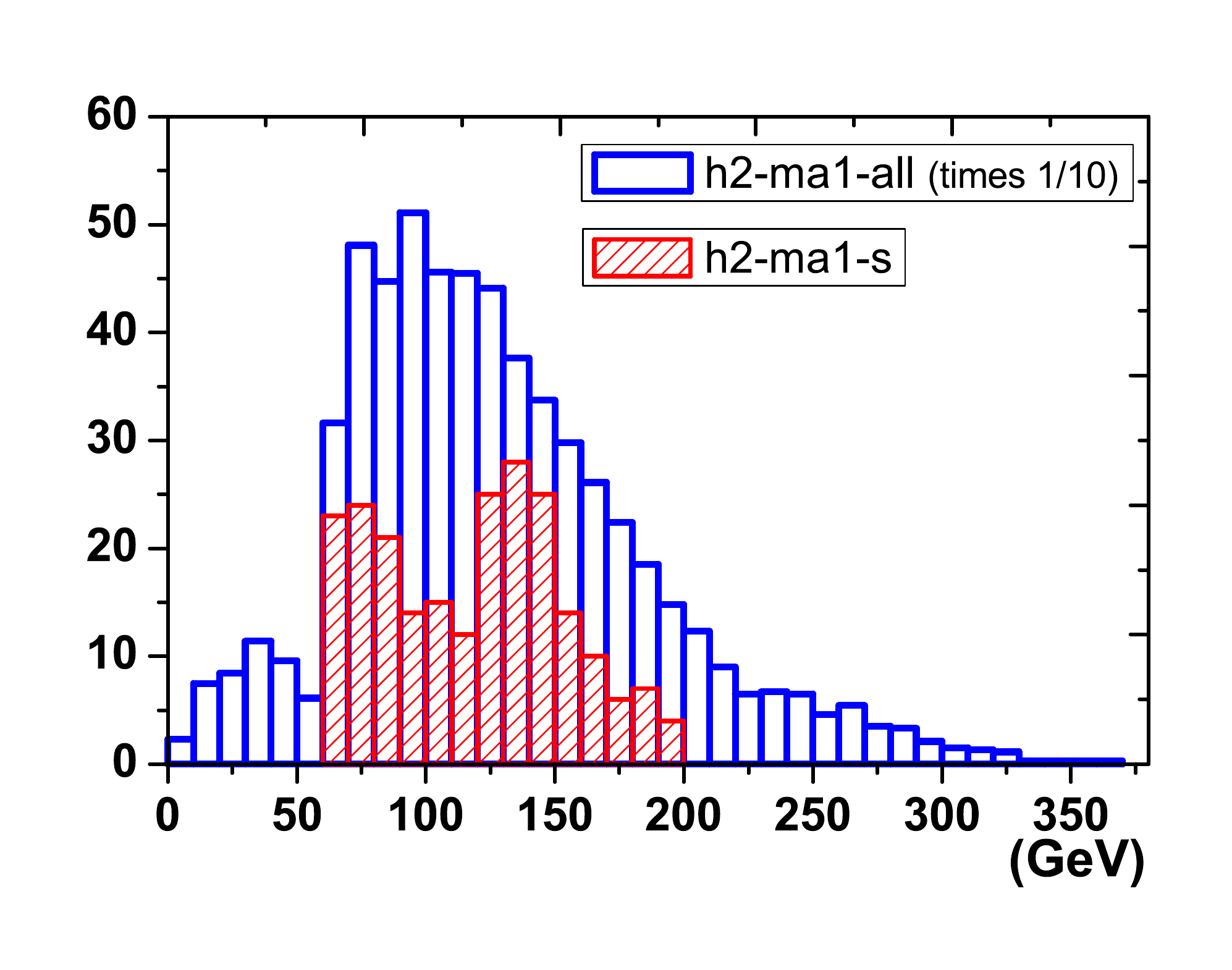}
\includegraphics[width=8cm]{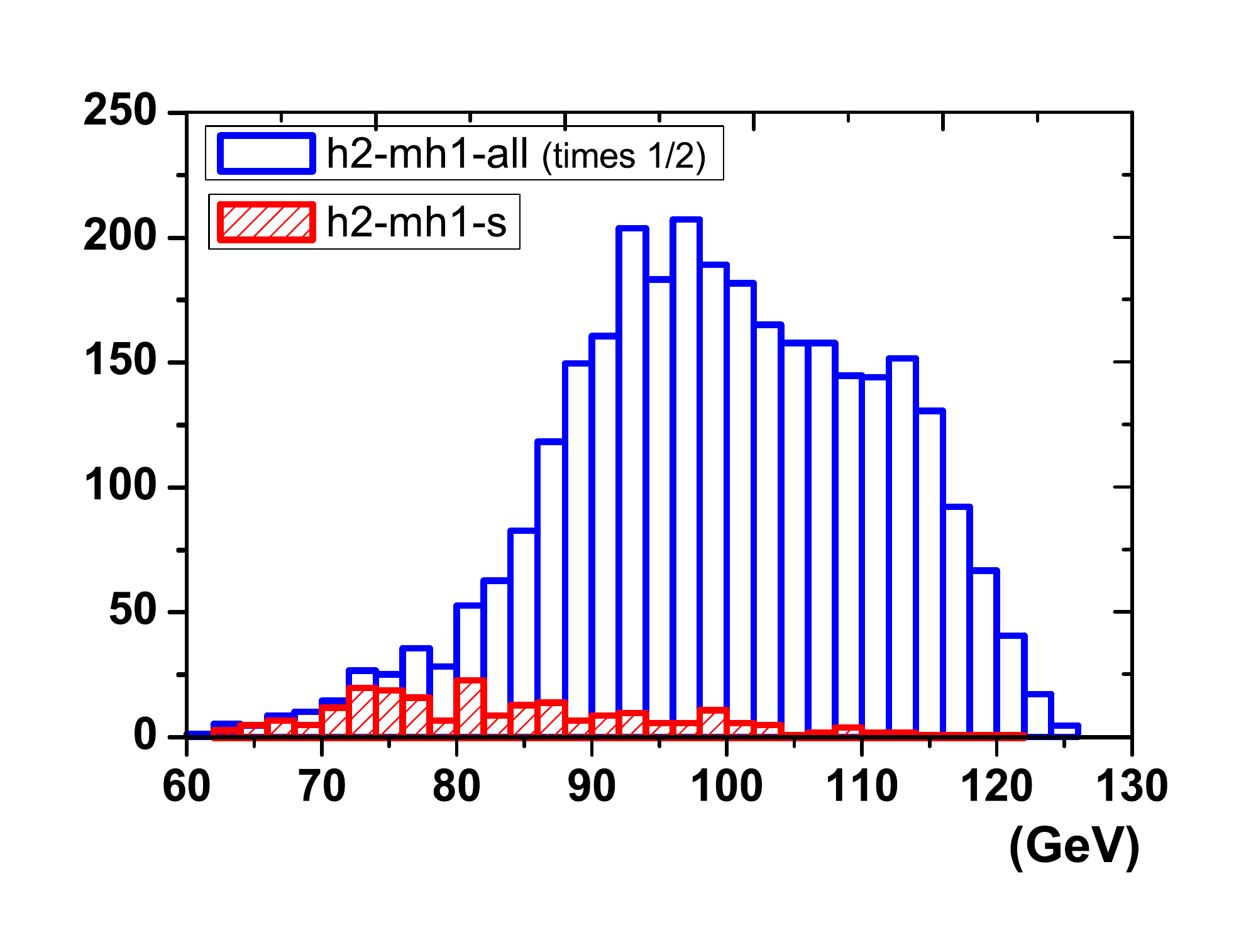}
\includegraphics[width=8cm]{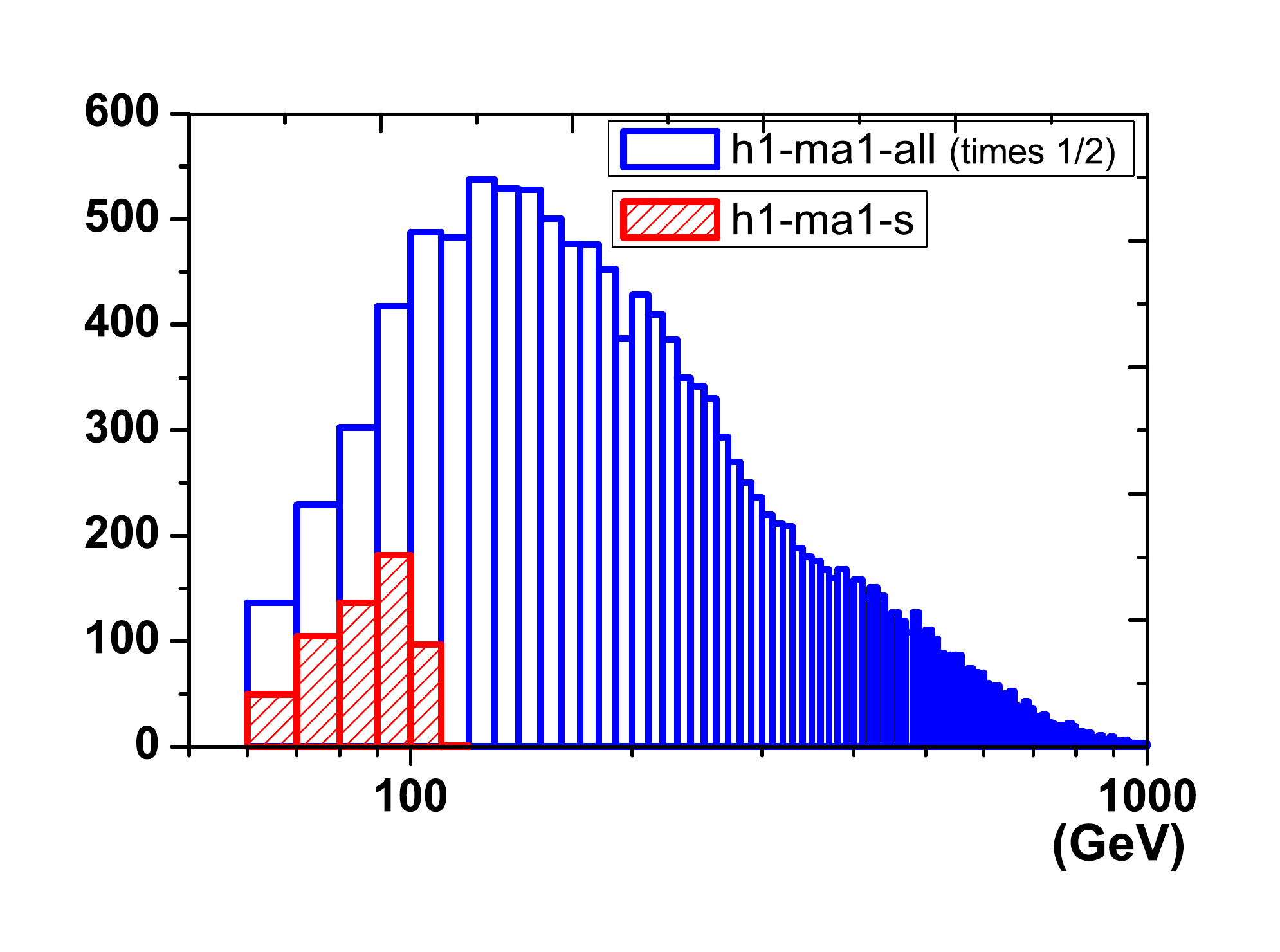}
\vspace*{-0.2cm}
\caption{Left panels: Histograms of $m_{a1}$ for physical points satisfying the CMS Higgs signal at $2\sigma$ level (blue-blank) and SFOEWPT (red-shaded) for Type-I samples in $H_{2}-$scenario (upper) and $H_{1}-$scenario (bottom) . Right panel: As in the left panels, histogram of $m_{a1}$ in $H_{2}-$scenario is plotted.}
\label{wpwfig1}
\end{figure}

\begin{figure}[t]
\includegraphics[width=8cm]{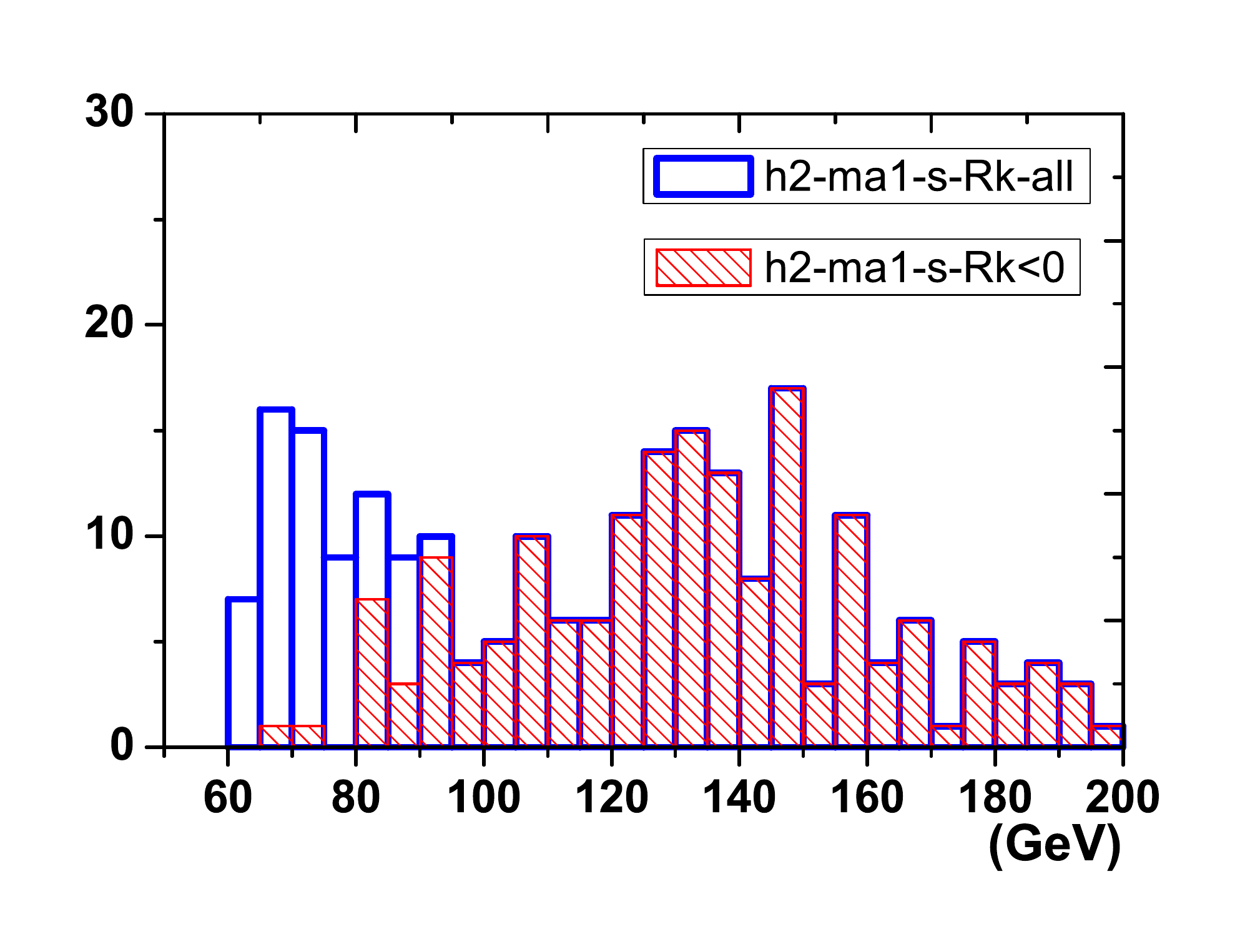}
\includegraphics[width=8cm]{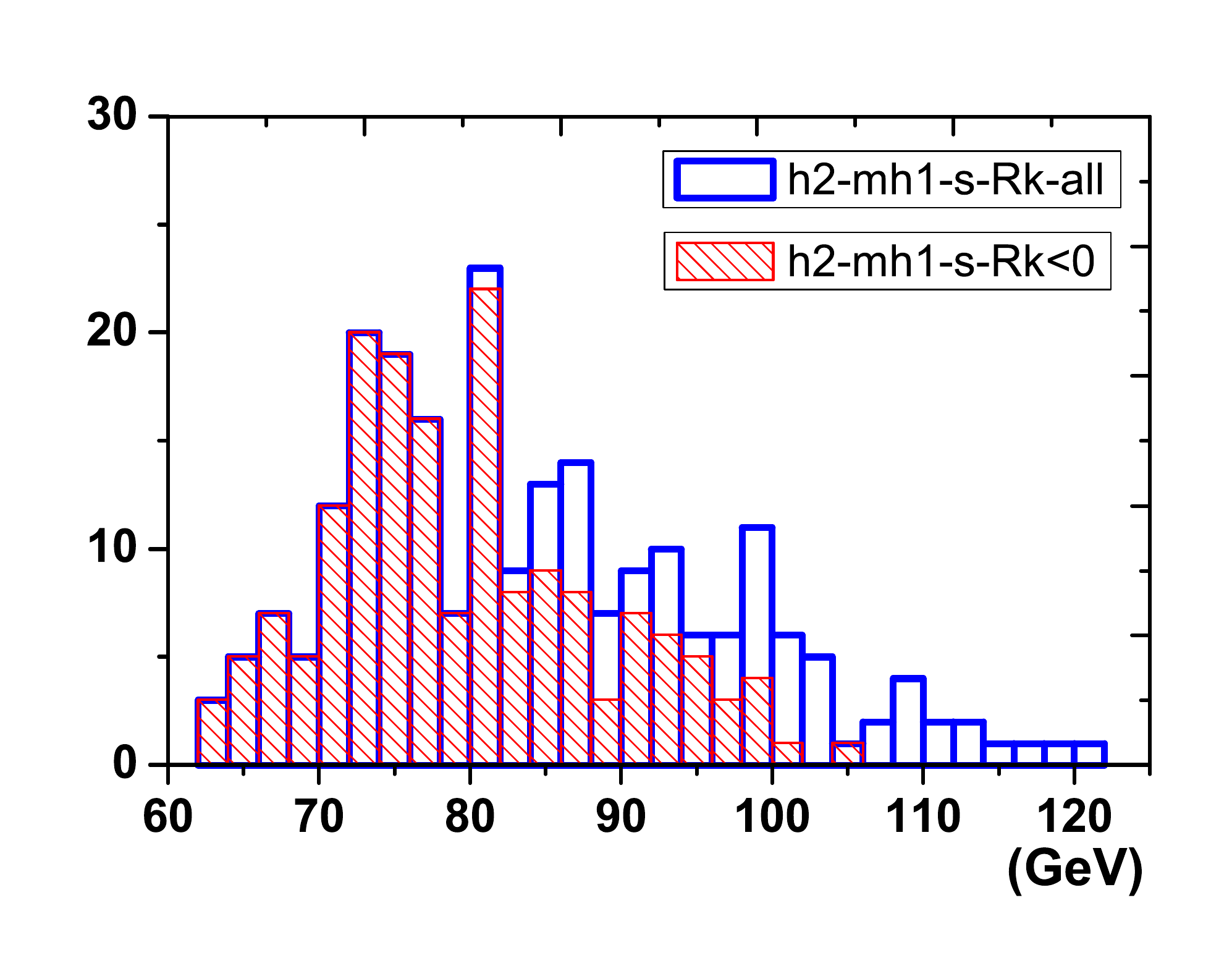}
\vspace*{-0.2cm}
\caption{Left panel: Histogram of $m_{a1}$ for physical SFOEWPT points with arbitrary $R_\kappa$'s (blue-blank) and points with $R_\kappa<0$ (red-shaded) for Type-I transition in $H_{2}-$scenario. Right panel: As in the left panel, histograms of $m_{a1}$ is plotted.}
\label{wpwfig2}
\end{figure}

The above speculation is confirmed by the histograms Fig. \ref{wpwfig1} and \ref{wpwfig2} in which we have imposed the CMS Higgs signal data on surviving samples in various channels including $h_{\rm SM} \to ZZ, WW, \gamma\gamma, bb,\tau\tau$ at $2\sigma$ level \cite{CMS_higgs_data} (ATALS constraints have a relative smaller number of points with the same distributions). In Fig. \ref{wpwfig1}, we compare the normal histograms for $m_{h1}$, $m_{a1}$ and those with SFOEWPT. We can see that the latter has more points concentrated on the light $m_{h1}$, $m_{a1}$ region. For those histograms which we have both $R_\kappa >0$ and $R_\kappa <0$, we further distinguish them in the histograms Fig. \ref{wpwfig2}. We can see that there is a very distinction that $R_\kappa > 0$ prefers small $m_{a1}$ while $R_\kappa < 0$ prefers small $m_{h1}$, which is coincident with our analysis above.

\subsection{Dark Matter Consideration}

If the lightest NMSSM neutralino is assumed to be the WIMP Dark Matter (DM) candidate, an important consideration is the DM relic density in the present epoch. Combining the PLANCK \cite{planck} and WMAP 9-year data \cite{wmap} and also including a 10\% theoretical uncertainty, the $2\sigma$ range of the WIMP DM relic density can be considered in the following range:
\begin{eqnarray}
0.091 \leq \Omega h^2 \leq 0.138
\label{relicdensity}
\end{eqnarray}
Since the bino and wino mass parameter $M_1,M_2$ have been fixed at 2 TeV in our analysis, the lightest neutralino basically consists of higgsino and/or singlino. From the neutralino mass matrix one can tell that, $\kappa$ can significantly affect the singlino component in the DM:
\begin{eqnarray}
{\cal M}_0 =
\left( \begin{array}{ccccc}
M_1 & 0 & -\frac{g_1 v_d}{\sqrt{2}} & \frac{g_1 v_u}{\sqrt{2}} & 0 \\
& M_2 & \frac{g_2 v_d}{\sqrt{2}} & -\frac{g_2 v_u}{\sqrt{2}} & 0 \\
& & 0 & -\mu & -\lambda v_u \\
& & & 0 & -\lambda v_d \\
& & & & 2 \kappa v_s
\end{array} \right)
\label{DMmassmatrix}
\end{eqnarray}
Since all squarks and sleptons have decoupled in our analysis, the DM can annihilate in the early universe only through the light Higgs bosons in the s-channel process or as the final states when the process is kinematically opened. For a highly higgsino-like DM, the coupling of the Higgs with the DM can be sizable and it is very easy to obtain a small relic density. We have checked that for each SFOEWPT scenario in our discussion, it is always possible to pick out several samples which can produce a relic density that does not overclose the universe, or even lies in the band shown in Eq.~(\ref{relicdensity}).

\section{Conclusion and discussion}

After the discovery of 126 GeV Higgs boson, the NMSSM is an attractive supersymmetric theory in virtue of its specific tree-level effect to enhance Higgs boson mass and allowing a more natural $\mu$ parameter. On top of that, the tree-level effect can readily enhance  the strength of EWPT $\gamma_c \gtrsim$1.0, which is required for a successful EWGB mechanism to generate the baryon asymmetry. In this article we have concentrated on studying SFOEWPT in the NMSSM, paying special attention on its relation with Higgs phenomenology. We have calculated EWPT strength $\gamma_c$ with the one-loop finite temperature effective potential and find that a larger $\gamma_c$ requires a smaller gap $\Delta V$. Then, in terms of the vacua structure and its evolution with temperature, we divide EWPT into three categories, Type-I, II and III along with two Higgs spectra patterns: $H_1-$scenario and $H_2-$scenario. We use our semi-analytical analysis as the intuitive understandings and then use our numerical results to confirm those understandings.

We have observed a dimensionless critical parameter $R_\kappa \equiv 4 \kappa v_s / A_\kappa$ which has demonstrated a clear correlation between the different types of SFOEWPT in NMSSM and the Higgs spectra as follows:
 \begin{itemize}
\item In $H_1-$scenario, the Type-I phase transition prefers $R_\kappa > 0 (\subset (5, 30))$ and much few points exists in the Type-III phase transition with small negative $R_\kappa$,
\item In $H_2-$ scenario, the Type-I phase transition has two distinct regions, in which either $0> R_\kappa > -1$ or $R_\kappa > 0 (\subset (2, 7))$. For the Type-III phase transition, much of them lies in the region $0 > R_\kappa > -4/3$.
  \end{itemize}
A SFOEWPT in general prefers a relatively light CP odd or even ($H_2-$scenario) Higgs mass. In particular, $R_\kappa > 0$ prefers small $m_{a1}$ while $R_\kappa < 0$ prefers small $m_{h1}$. 

The current classification on the EWPT patterns and the Higgs spectra has a great importance to guild us to the next step of understanding the nature of EWPT. In particular, the Higgs spectra in SFOEWPT in the NMSSM, or even broadly, in some SM / 2HDM + singlet models, prefers either a light CP odd or even Higgs with their mass slightly more than $m_{h_{SM}}/2$. We hope that our work can help bring more attention on the observation of those (60, 100) GeV light Higgs and works along this direction will be presented in the future \footnote{The extremely light Higgs search from the 126 GeV Higgs exotic decay in the PQ limit \cite{Draper:2010ew} is studied in Ref. \cite{Huang:2013ima}. }.

\section{Acknowledge}

We thank the National Natural Science Foundation of China under grant Nos. 11275245, 10821504 and 11135003.

\clearpage

\appendix
\section{Minimum in the singlet subspace\label{appa-min}}
\label{apx:prop}
For convenience, we present the minimum structure in the singlet subspace here. The potential has a form
\begin{align}\label{}
  V(S)=\frac{b_2}{2}S^2+\frac{b_3}{3}S^3+\frac{b_4}{4}S^4.
\end{align}
The minimum condition $\partial V/\partial S=0$ has following solutions:
\begin{eqnarray}
  S &=& 0 \\
  S_{\pm} &=& \frac{-b_3\pm \sqrt{b_3^2-4b_2b_4}}{2b_4}
\end{eqnarray}
The latter is physical as long as $\Delta\equiv b_3^2-4b_2b_4 >0$, i.e, $x\equiv b_2b_4/(2b_3^2)\leq 1/8$.

The minimum structures are listed in Table~\ref{min-stru}. Sometimes, it is useful to express the potential at nontrivial minimum as
\begin{align}
  V_{\rm min}(S_{\pm}) &=\frac{b_3}{4}(\frac{b_2}{b_3}+\frac{S_{\pm}}{3})S_{\pm}^2 =-\frac{b_3^4}{96b_4^3}f(x)
\end{align}
where $ f(x)=4+4\sqrt{1-8x}-48x-32x\sqrt{1-8x}+96x^2$, which decreases monotonically from one to zero as x increares from zero to $2/9$. Obviously, $V_{\rm min}=0$ for $x>2/9$.

\begin{table}[t]
  \centering
  \begin{tabular}{|c|c|c|c|c|}
\hline
\multicolumn{4}{|c|}{The range of parameters} & minimum\tabularnewline
\hline
\multicolumn{4}{|c|}{$x>\frac{1}{8}$} & 0\tabularnewline
\hline
\multirow{5}{*}{$x<\frac{1}{8}$}  & \multirow{3}{*}{$b_{2}>0$} & \multicolumn{2}{c|}{$\frac{1}{8}>x>\frac{1}{9}$} & 0\tabularnewline
\cline{3-5}
 &  & \multirow{2}{*}{$x<\frac{1}{9}$} & \multicolumn{1}{c|}{$b_{3}<0$} & $S_{+}$\tabularnewline
\cline{4-5}
 &  &  & $b_{3}>0$ & $S_{-}$ \tabularnewline
\cline{2-5}
& \multirow{2}{*}{$b_{2}<0$} & \multicolumn{2}{c|}{$b_{3}<0$} & $S_{+}$\tabularnewline
\cline{3-5}
 &  & \multicolumn{2}{c|}{$b_{3}>0$} & $S_{-}$\tabularnewline
\cline{2-5}
\hline
\end{tabular}
  \caption{Minimum structures in singlet direction}\label{min-stru}
\end{table}

\end{document}